\documentclass[conference]{IEEEtran}

\usepackage[T1]{tipa}
\usepackage[tt=false, type1=true]{libertine}
\usepackage[libertine]{newtxmath}
\usepackage[shortcuts]{extdash}
\usepackage{seqsplit}

\usepackage{hyperref}

\usepackage{stackengine}
\usepackage{rotating}
\usepackage[clock]{ifsym}
\usepackage{fontawesome5}

\usepackage{tikz,pgfplots}
\usetikzlibrary{arrows.meta}

\pgfplotsset{every axis/.append style={font=\sffamily\scriptsize}}
\usepackage{float}
\usepackage{tipa}
\usepackage{graphicx}
\usepackage{balance}
\usepackage{stfloats}
\usepackage[caption=false,font=footnotesize,labelfont=sf,textfont=sf]{subfig}
\usepackage{booktabs}
\usepackage[symbol]{footmisc}
\usepackage{makecell}
\usepackage{mdframed}
\usepackage{listings}
\usepackage{bm}
\usepackage{enumitem}
\usepackage{multirow}
\usepackage{nicefrac}
\usepackage{amsmath}

\usepackage{amssymb}
\usepackage{booktabs}
\usepackage{todonotes}
\usepackage{arydshln}
\usepackage{booktabs,xcolor,siunitx}
\usepackage{oplotsymbl}

\usepackage{soul}
\usepackage{xcolor}
\usepackage{tcolorbox}
\definecolor{lightgrey}{HTML}{D3D3D3}
\definecolor{commentblue}{HTML}{155196}
\definecolor{emphgreen}{HTML}{4F6900}
\definecolor{emphpink}{HTML}{b5283c}
\definecolor{coloronlybulk}{HTML}{5A1F72}
\definecolor{colorsat}{HTML}{008000}
\definecolor{colornvhpc}{HTML}{B3E680}
\definecolor{colorgcc}{HTML}{D3CF8F}
\definecolor{colorclang}{HTML}{80E5E7}

\definecolor{lightgrey2}{HTML}{E8E8E8}
\newcommand{\ctext}[1]{{#1}}

\def\BibTeX{{\rm B\kern-.05em{\sc i\kern-.025em b}\kern-.08emT\kern-.1667em\lower.7ex\hbox{E}\kern-.125emX}}
    
\newtcbox{\mybox}[1][black]{on line,
  height=1.2em,
  colback=#1, colframe=#1, boxsep=0pt, boxrule=0pt, size=small, arc=1mm}

\begin{document}

\title{ACC Saturator: Automatic Kernel Optimization for Directive-Based GPU Code}

\author{\IEEEauthorblockN{Kazuaki Matsumura}
\IEEEauthorblockA{\textit{Barcelona Supercomputing Center (BSC)} \\
kmatsumura@nvidia.com\vspace*{-1.0cm}}
\and
\IEEEauthorblockN{Simon Garcia De Gonzalo}
\IEEEauthorblockA{\textit{Sandia National Laboratories}\\
simgarc@sandia.gov\vspace*{-1.0cm}}
\and
\IEEEauthorblockN{Antonio J. Peña}
\IEEEauthorblockA{\textit{Barcelona Supercomputing Center (BSC)} \\
antonio.pena@bsc.es\vspace*{-1.0cm}}}

\setlength{\textfloatsep}{10pt}
\setlength{\dbltextfloatsep}{10pt}
\setlength{\abovedisplayskip}{0pt}
\setlength{\belowdisplayskip}{6pt}
\setlength{\abovedisplayshortskip}{0pt}
\setlength{\belowdisplayshortskip}{6pt}

\maketitle
\begin{abstract}
Automatic code optimization is a complex process that typically involves the application of multiple discrete algorithms that modify the program structure irreversibly. However, the design of these algorithms is often monolithic, and they require repetitive implementation to perform similar analyses due to the lack of cooperation. To address this issue, modern optimization techniques, such as equality saturation, allow for exhaustive term rewriting at various levels of inputs, thereby simplifying compiler design.

In this paper, we propose equality saturation to optimize sequential codes utilized in directive\-/based programming for GPUs. Our approach realizes less computation, less memory access, and high memory throughput simultaneously. Our fully\-/automated framework constructs single\-/assignment forms from inputs to be entirely rewritten while keeping dependencies and extracts optimal cases. Through practical benchmarks, we demonstrate a significant performance improvement on several compilers. Furthermore, we highlight the advantages of computational reordering and emphasize the significance of memory\-/access order for modern GPUs.
\vspace{4pt}
\vspace{0.1cm}
\end{abstract}

\begin{IEEEkeywords}
Compiler, Code Generation, GPUs, Program Optimization, Directive-Based Programming
\end{IEEEkeywords}

\section{Introduction}\label{sec:introduction}

Over the last decade, the increasing popularity of computational accelerators in high-performance computing (HPC) has significantly impacted the programming landscape. The emergence of complex, unique, yet highly efficient architectures has posed a challenge for scientists, who must exert additional effort to take advantage of these architectures. Among the fastest computers, Graphics Processing Units (GPUs) are the most pervasive accelerators~\cite{top500} and are inherently parallel, requiring software to expose concurrent calculations. Traditional optimization techniques may not provide performance benefits in such scenarios. Furthermore, with the approach of the end of Moore's Law~\cite{Moore1998}, accelerators have become increasingly domain-specific, and compilers have become essential for the automatic utilization of supercomputing resources.

Although coding with low-level languages is typically the most profitable approach, utilizing them often involves complex data dependencies. As a result, adapting vendor\-/specific languages like CUDA~\cite{cuda} and OpenCL~\cite{opencl} requires extensive knowledge of the application and creates separate code entities just for computation offloading, which increases programming costs. To address this challenge, current GPU accelerators support programming with abstract models. OpenACC~\cite{openacc} and OpenMP~\cite{openmp} have emerged as popular models that alleviate the difficulty of accelerator use by providing code directives to existing languages. These directives allow users to specify compute\-/intensive parts of the original code, which compilers may automatically translate into accelerator code. At runtime, offloading is realized without any user intervention.

Directive-based code typically maintains a general programming style and remains unspecialized until compilation. It is the responsibility of compilers to handle the computation, the redundancy of memory accesses, and the order of those while considering the nature of accelerators. However, code rewriting while preserving semantics may be challenging. Current compilers apply optimization for each metric sequentially, which limits opportunities for improved performance.
In particular, user-specified directives fix loop structures, so little work has been done on optimising code under directives other than relying on compiler techniques used for sequential programs~\cite{barua, jacc}.
As a result, more complex approaches are needed to improve the performance of directive-based code.

This paper proposes the use of a modern optimization technique, {\it
  equality saturation}~\cite{equality}, to fine-tune directive\-/based
GPU code. Our tool, {\bf ACC Saturator}\footnote[4]{The artifact is available at \url{https://github.com/khaki3/acc-saturator}.}, achieves less computation,
fewer memory accesses, and high throughput at the same time, while
easily integrating into the compilation of both OpenACC and OpenMP
compilers. ACC Saturator performs kernel optimization by passing
programs through an e-graph, a graph structure for equality
saturation. Unlike other optimization methods, our approach neither
transforms the abstract syntax trees nor it changes directives. Despite this, we attain significant performance improvements of up to 2.23x with the NVHPC compiler and 5.08x with GCC. We present a detailed performance analysis that highlights the benefits of each optimization on the state-of-the-art GPU architecture, NVIDIA Tesla A100. Our contributions are four-fold:
\begin{enumerate}[topsep=0.8em, leftmargin=0.2in,rightmargin=0.05in, itemsep=6pt]
\item We develop a fully automated OpenACC/OpenMP framework for equality saturation.

\item We combine static single-assignment form with e-graph to optimize directive-based code.

\item We demonstrate that our approach provides significant performance opportunities for GPUs.

\item We provide a detailed analysis of kernel performance and the
  effectiveness of optimization techniques using the latest NVIDIA GPU
  architecture for HPC.
\end{enumerate}

The rest of the paper is organized as follows.
Section~\ref{sec:background} provides background knowledge on
GPUs, directive-based programming, code optimization, and equality saturation.
Section~\ref{sec:overview} presents a high-level overview of our work.
Sections~\ref{sec:rep} and~\ref{sec:sat} explain
our approach to represent general sequential code for equality saturation
and optimizations.
Section~\ref{sec:codegen} describes our novel code generation technique
from e-graphs towards high\-/throughput GPU execution.
In Section~\ref{sec:method}, we describe our experimental methodology.
Section~\ref{sec:evaluation} presents the results of our evaluation
along with in-depth analysis on each technique.
Section~\ref{sec:related} discusses related work.
Section~\ref{sec:conclusion} provides concluding remarks.

\section{Background}\label{sec:background}
This section provides an overview of GPUs, directive-based programming, and equality saturation.

\subsection{GPUs}
Parallel computing is an assemblage of sequential execution, with means of intercommunication and mutual effects on efficiency.
Such architectures accommodate simultaneous operations on multiple different data in uniform processors that are adjacent to each other.
As an instance, a Graphics Processing Unit (GPU) runs hundreds of thousands of {\it threads} at the same time on around a hundred of {\it streaming multiprocessors (SM)}.
One of the most recent GPUs, NVIDIA Tesla A100, features 108 SMs and each SM contains 32 FP64 cores along with multiple LD/ST units~\cite{a100}.

The primary benefit of GPU execution stems from overlapping utilization of both computational and memory units. Although GPUs provide high memory bandwidth, the global memory access latency is also higher than that of CPUs~\cite{dissecting-volta, dissecting-turing, cpu-latency}. Therefore, optimal throughput may be attained by covering memory requests with computational execution and hiding the latency of data movement. Additionally, the order of memory accesses is essential due to the memory hierarchy, because the distance among accesses by neighboring threads often results in limited bandwidth~\cite{gpu-pad}.

\subsection{Directive-Based Programming}

Decomposing sequential execution into parallel entities is
challenging. Routine calls often create data dependencies across
multiple files, hindering the extraction of compute-intensive
code. Even with extracted code, loops often possess intricate access
patterns on different structures that challenge automatic
parallelization for efficiency. Directive-based programming models,
such as OpenACC~\cite{openacc} and OpenMP~\cite{openmp}, alleviate the
burden of introducing additional code for hardware acceleration. Both
OpenACC and OpenMP provide code directives on the {\it de facto} standard languages for scientific applications (C/C++/ Fortran) to specify offloading parts as {\it compute kernels} with the declaration of explicit parallelism, enabling seamless automatic accelerator use.

Listing~\ref{lst:acc} provides an example of OpenACC code that utilizes code directives to specify the parallelism of loops.
The \texttt{\color{emphgreen}\bfseries kernels} directive guards a sequence of kernels,
while the {\color{emphgreen}\bfseries loop} directive explicitly sets the parallelism of a loop.
Users may also specify a single kernel region by using the \texttt{parallel} directive (not shown in the listing).
OpenACC's parallelism consists of three levels of abstraction: gang,
worker and vector (in coarse to fine order). For instance, on the NVHPC compiler, 
both the top \texttt{i} and the middle \texttt{j} loops feature gang parallelism with the degree of the trip count \texttt{cy} and 16, respectively, and these are distributed over thread-blocks on GPUs.
The middle \texttt{j} loop exposes additional vector parallelism and launches 256 threads to execute the inner statements. The blue comments show an equivalent form in OpenMP, which does not allow the reuse of parallelism across nested loops.
Loop parallelism and data transfers may be implicit, and the compiler automatically solves data dependencies and sets optimal parameters. The NVHPC compiler generates embarrassingly parallel code for GPUs from directive code. On the other hand, GCC utilizes a principal-agent model for both OpenACC and OpenMP, and Clang follows the same approach for OpenMP~\cite{ptxasw}.

\begin{figure}[t]
\vspace{4pt}
\begin{lstlisting}[caption={Matrix multiplication kernel in C and
    OpenACC (OpenMP equivalent commented)}, label={lst:acc}]
_*#pragma acc kernels loop independent*_
//#pragma omp target teams distribute
[*for*] (int i = 0; i < cy; i++) {
_*#pragma acc loop independent gang(16) vector(256)*_
//#pragma omp parallel for simd
  [*for*] (int j = 0; j < cx; j++) {
    double tmp = 0.f;
    [*for*] (int l = 0; l < ax; l++)
      tmp += a[i][l] * b[l][j];
    r[i][j] = alpha * tmp + beta * c[i][j];
}}
\end{lstlisting}
\vspace{-10pt}
\end{figure}

\subsection{Limits of Annotation Optimization}
\label{sec:limit}

GPU architectures require substantial parallelism to achieve high efficiency. Therefore, configuring parallelism appropriately based on the execution model is crucial. In directive-based programming, parallelism can be controlled through clauses like 'gang', 'worker', and 'vector'. These annotations significantly influence the mapping of code to hardware. For instance, the work by  Lambert {\it et al.}~\cite{ccamp}. illustrates that vector parallelism can lead to significant performance degradation by underutilizing thread-blocks. Conversely, Gong {\it et al.}~\cite{gong2016nekbone}. demonstrate that specifying gang and vector parallelisms tailored to each compiler can lead to performance improvements over the compiler's default settings. This suggests that application developers should fine-tune these directives based on actual hardware and software environments.

Moreover, once the parallelism annotations are set, any structural changes to the application may introduce performance bottlenecks. Techniques such as loop unrolling increase register usage, while loop folding may constrain parallelism, both of which reduce SM occupancy. Consequently, compilers must develop optimization strategies that enhance application performance not by altering the structural integrity but by refining the computational sequence.  Our proposed {\bf ACC Saturator} is designed to effectively accomplish this.

\subsection{Equality Saturation}

Compiler techniques have been critical in utilizing the best resources available for program execution on the target architecture for the last half-century~\cite{dragon, hpc4pc}.
Various features of hardware, including registers, cache systems, pipelines, and parallelization, rely on the compiler's efforts for performance, enabling target-specific optimization under resource constraints.
However, compilers tend to miss a holistic view of performance opportunities and require additional discrete efforts for each discrete optimization mechanism~\cite{rawatgpu}.
{\it Equality saturation}, a state\-/of\-/the\-/art technique for compilers~\cite{equality}, defines a set of rewriting rules over the accumulation of equal expressions of target code. The rewriting continues until the expression gets {\it saturated}, finding no other forms, or hits a time or size limitation, attaining optimal codes based on a cost model that considers the entire computation.

A dedicated graph structure called {\it e-graph} accepts the accumulation of equal expressions, while sharing redundant code blocks over expressions.
The graph structure consists of {\it e-classes}, groups of equal {\it e-nodes}.
One e-node may point to e-classes as its subexpressions, and the accumulation works by extending or merging e-classes in accordance with rewriting rules, while preserving the relationship between parental e-nodes and child e-classes~\cite{egg, evm}.
To extract optimal solutions, each e-class selects one contained e-node with a minimum cost, while optionally counting common expressions only once and completing common subexpression elimination (CSE).

\section{Methodology Overview}\label{sec:overview}

\begin{figure*}[t]
  \vspace*{0.15cm}
  \centering
  \noindent\makebox[\textwidth]{
  \includegraphics[width=\textwidth]{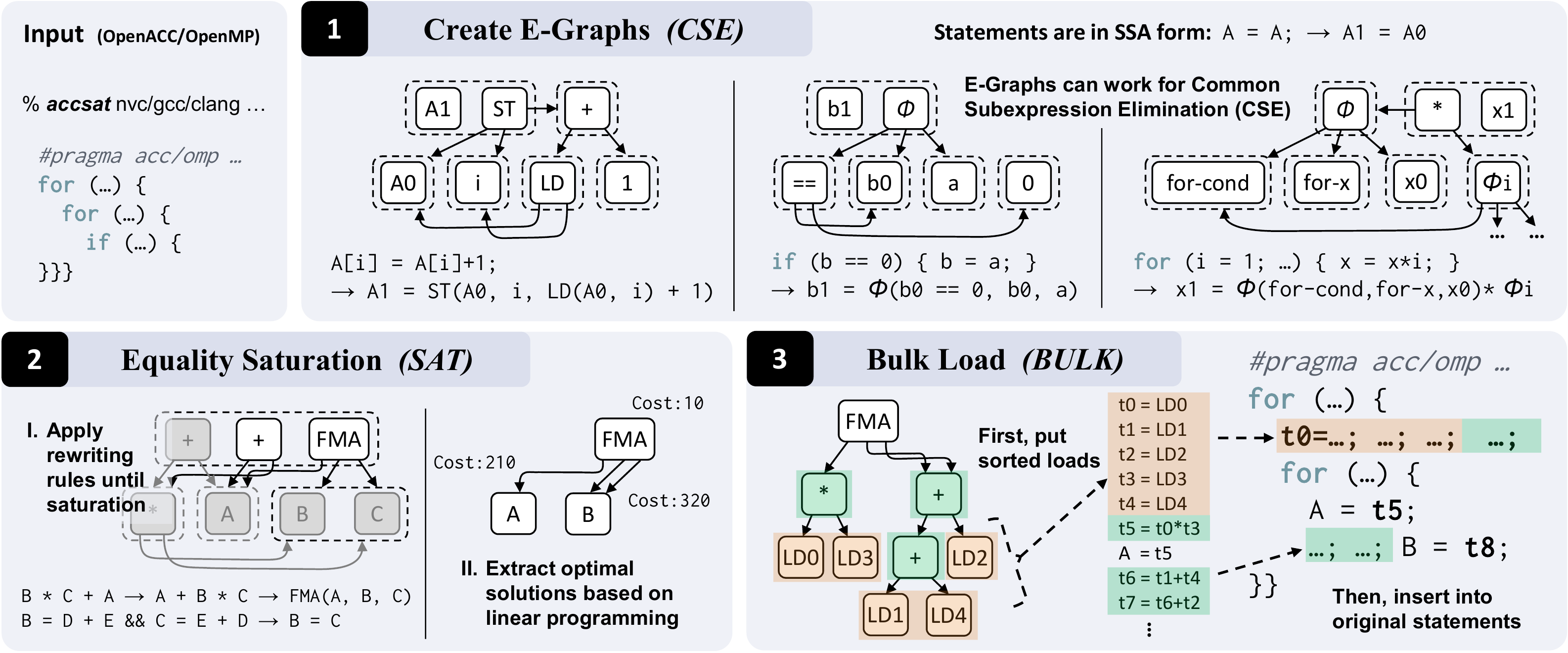}
  }
  \vspace*{-0.6cm}
  \caption{Overview of ACC Saturator}
  \label{fig:overview}
  \vspace*{-0.0cm}
\end{figure*}

ACC Saturator is the implementation of our proposal of equality saturation for directive-based code. We provide a convenient command\-/line tool that wraps normal C-compiler invocation and replaces the original inputs with saturated codes.

Figure~\ref{fig:overview} provides an overview of our work.
ACC Saturator optimizes the sequential parts of parallel loops by packing and unpacking e-graphs for extracted expressions while preserving original code structures.
Array references, branching, loops, function calls, and member references are supported under a dataflow abstraction.
Once e-graphs have the inputs, we run equality saturation with an arbitrary set of rewriting rules and succeedingly select optimal solutions featuring minimal total costs according to our model.
With new expressions, we update user code while arranging the order of
computation and continue to the compiler invocation, since generated code is compatible with NVHPC, GCC, and Clang. 

The e-graph operation by ACC Saturator consists of three phases.
First, \mybox{\raisebox{-0.5em}[-0.0em][-0.0em]{\color{-.}\bfseries 1}}
we build a static single-assignment form (SSA) from the input code to create initial e-graphs (Section~\ref{sec:rep}).
E-graphs accept our SSA representation holding dependencies, conditions, and iterations over C-style operations.
Second, \mybox{\raisebox{-0.5em}[-0.0em][-0.0em]{\color{-.}\bfseries 2}}
we define rewriting rules and a cost model for performance improvement (Section~\ref{sec:sat}). Our tool runs equality saturation and extracts optimal solutions under time and size limitations.
Last, \mybox{\raisebox{-0.5em}[-0.0em][-0.0em]{\color{-.}\bfseries 3}} we generate output code (Section~\ref{sec:codegen}).
As a result, user kernels are reduced to minimum computation with a new order.

\section{Program Representation}\label{sec:rep}

During compiler optimizations, the program semantics and behavior are preserved while improving its performance. Maintaining the order of data accesses is especially important to ensure reproducibility of results.
To express clear dependencies among statements, program analysis often relies on
{\it static single-assignment form (SSA)}, which allows only one definition per variable~\cite{appel_1997}.
Early work on equality saturation~\cite{equality} used an SSA-based graph structure to represent and rewrite an entire program.
For directive\-/based code, however, user-specified parallelism
significantly affects performance~\cite{autotuningacc}, so compilers
are limited to respect users' decisions.

Our proposed methodology optimizes OpenACC/OpenMP code while
preserving a connection between e-graphs and original code. We track
optimal expressions in e-graphs using SSA variables and insert
optimized code into programs with the same structures and directives
as input. The rest of this section covers the conversion process from directive\-/based code to e-graphs and explains the solution finding process based on SSA. By doing so, we are able to improve the performance of parallel code without changing its semantics or behavior.

ACC Saturator is a more practical compiler optimization framework than previous work~\cite{equality} because it maintains code structures, is capable of restoring the original state, and is applicable to standard languages such as C and Fortran. ACC Saturator allows equality saturation over directive-based code, upholding the importance of structure information which previous work disregarded. Reusing the code style that the input provides, ACC Saturator can support the user's intended target architecture by rewriting rules that lead to a new order of computation. Each rule is simple, but by combining many rules, we discover better solutions and attain speedups over the state-of-the-art industry compilers.

\subsection{E-Graph Creation}
Both OpenACC and OpenMP contain sequential parts within the innermost parallel loops that may be executed independently across multiple threads and thus optimized for more efficient execution or reduced computation. To attain this, we create an e-graph for each innermost parallel loop as follows:
First, we introduce conditional $\phi$ nodes~\cite{gatessa} to represent control structures such as {\texttt if} and {\texttt for}, while merging data flows.
Second, we assign an ID to each variable/array assignment or $\phi$.
Third, we update each variable/array load to refer to the latest ID along its data flow.
Last, for each assignment or $\phi$, we assign both the ID and the expression to the same e-class.

Examples of e-graphs for store/load, {\texttt if}, and {\texttt for}
operations are illustrated in
\mybox{\raisebox{-0.5em}[-0.0em][-0.0em]{\color{-.}\bfseries 1}} at Figure~\ref{fig:overview}, with some IDs omitted for simplicity. The e-nodes are depicted as white blocks, and the e-classes as dotted boxes, while the black arrows represent parent-child relationships among them.
As shown in each example, multiple references to the same variables are now directed toward common e-classes, thereby reducing redundancy.
For each updated variable, $\phi$ is created within {\texttt if} and
{\texttt for} structures with conditions that may be concrete or abstract. Our tool supports C-style operations, such as function calls, pointers, and member references, within the same SSA framework. To optimize the code, ACC Saturator leverages our e-graph representation with rewriting rules and cost models, as detailed in Section~\ref{sec:sat}.

\subsection{Code Selection}\label{sec:codeselect}

To select optimal codes in the e-graph, we aim to find an equal expression that corresponds to a set of all assignments. We extract the lowest-cost expression that contains all the e-classes of assignments based on a sequence of IDs. The total cost is calculated as the sum of the cost of each e-class, with common e-classes being counted only once. To attain this, we use linear programming techniques~\cite{cbc}. 

Each assignment is updated based on its ID and the extracted
expression. To enable reuse across assignments, the values of common
expressions are stored in temporary variables. ACC Saturator allows
for customizing the generation order of these temporary variables (Section~\ref{sec:codegen}).

\section{Optimization with Saturation}\label{sec:sat}

In this section, we outline our methodology for enhancing OpenACC/OpenMP code performance through equality saturation. Our approach is based on a meticulously developed set of rewriting rules paired with a cost model designed to optimize computation and memory access efficiencies. As highlighted in Section~\ref{sec:limit}, structural modifications in the code may cause significant performance variations. Additionally, the architectural specifics in directive-based programming often remain undisclosed. To address these challenges, our primary strategy involves optimizing the sequence and volume of computations, providing a generalized solution to accelerate application performance.

\subsection{Rewriting Rules}\label{sec:rule}

Reordering computations allows programs to explore different
optimization possibilities. Rather than selecting a single
alternative, compilers aim to choose the most profitable candidate
according to their objective. Equality saturation defers this decision
and instead accumulates equal expressions using multiple rewriting
rules. With this approach, ACC Saturator can derive efficient
operations and facilitate the reuse of expressions simultaneously, as
shown in \mybox{\raisebox{-0.5em}[-0.0em][-0.0em]{\color{-.}\bfseries
    2}} at Figure~\ref{fig:overview}.

We apply two sets of rewriting rules to our equality saturation process. The first set introduces fused multiply\-/adds (FMA) operations, which can improve code generation and resource utilization on GPUs~\cite{rawatgpu}. Table~\ref{tab:rule} lists the minimum set of rules we use. When we encounter expressions that match the FMA pattern, we add the corresponding operations to the e-classes of matched expressions. The second set of rules benefits from the commutative and associative properties of the plus and multiply operators to reorder computation. This can enable common subexpression elimination and produce new FMA operations. We also incorporate constant folding of arithmetic operations with integer and floating-point numbers.

While ACC Saturator can rewrite subtraction, division, memory access order, conditional expressions, and iterations, these rules can increase the size of e-graphs and lead to slow extraction of optimal solutions in real-time. Therefore, we restrict the tool to only use the set of rules mentioned earlier for efficient performance.

\begin{table}[h]
\vspace{-8pt}
\center
\caption{ACC Saturator's rewriting rules}
\label{tab:rule}
\center
\begin{tabular}{c|lcl}
Name & \multicolumn{1}{c}{Pattern} & & \multicolumn{1}{c}{Result} \\\hline
\textbf{FMA1} & \texttt{A + B * C} & $\rightarrow$ & \texttt{FMA(A, B, C)} \\
\textbf{FMA2} & \texttt{A - B * C} & $\rightarrow$ & \texttt{FMA(A, -B, C)} \\
\textbf{FMA3} & \texttt{B * C - A} & $\rightarrow$ & \texttt{FMA(-A, B, C)} \\
\textbf{COMM-ADD} & \texttt{A + B} & $\rightarrow$ & \texttt{B + A} \\
\textbf{COMM-MUL} & \texttt{A * B} & $\rightarrow$ & \texttt{B * A} \\
\textbf{ASSOC-ADD1} & \texttt{A + (B + C)} & $\rightarrow$ & \texttt{(A + B) + C}\\
\textbf{ASSOC-ADD2} & \texttt{(A + B) + C} & $\rightarrow$ & \texttt{A + (B + C)}\\
\textbf{ASSOC-MUL1} & \texttt{A * (B * C)} & $\rightarrow$ & \texttt{(A * B) * C}\\
\textbf{ASSOC-MUL2} & \texttt{(A * B) * C} & $\rightarrow$ & \texttt{A * (B * C)}\\
\end{tabular}
\vspace{-4pt}
\end{table}

\subsection{Cost Model}\label{sec:cost}

GPUs are complex systems that require careful analysis to accurately
predict their efficiency~\cite{accelsim,wuestim}. While applications
running on GPUs often face memory-bound
limitations~\cite{gpubottleneck}, it is not just the number of memory accesses that affects overall bandwidth. Factors such as on-chip resource utilization, processor occupancy, thread/grid-level parallelism across multiple memory layers, and instruction-level parallelism (ILP) all play a role, and improving one metric often comes at the expense of another. In ACC Saturator, our focus is on reducing memory access and computation by utilizing registers for common expressions, while maintaining ILP as described in Section~\ref{sec:codegen}.

Our cost model is simple: constant numbers pose no cost, each input variable or $\phi$ counts as 1, all computational operations except division and modular arithmetic count as 10, and each memory access, division, modular arithmetic, or function call counts as 100. The assigned costs are based on empirical testing and aim to reflect the relative cost of different operations. We acknowledge that future research could refine the cost values.

\begin{figure}[t]
\vspace{4pt}
\begin{lstlisting}[caption={One of kernels in NPB-BT's z\_solve.c}, label={lst:zsolve}]
_*#pragma acc parallel loop gang num_gangs(ksize-1)\
                  num_workers(4) vector_length(32)*_
[*for*] (k = 1; k <= ksize-1; k++) {
_*#pragma acc loop worker*_
  [*for*] (i = 1; i <= gp02; i++) {
_*#pragma acc loop vector*_
    [*for*] (j = 1; j <= gp12; j++) {
      temp1 = dt * tz1; temp2 = dt * tz2;
      lhsZ[0][0][AA][k][i][j] =
   - temp2 * fjacZ[0][0][k-1][i][j]
   - temp1 * njacZ[0][0][k-1][i][j] - temp1 * dz1;
      // ... Similar 74 statements continue
}}}
\end{lstlisting}
\begin{lstlisting}[caption={Generated Code of ACC Saturator (formatted)}, label={lst:zsolvegen}]
_*#pragma acc parallel loop gang num_gangs(ksize-1)\
                  num_workers(4) vector_length(32)*_
[*for*] (k = 1; k <= ksize-1; k++) {
_*#pragma acc loop worker*_
  [*for*] (i = 1; i <= gp02; i++) {
_*#pragma acc loop vector*_
    [*for*] (j = 1; j <= gp12; j++) {
      double _v277, _v274, _v3 /* ... */;
      _v277 = njacZ[0][0][k][i][j];
      _v274 = njacZ[0][1][k][i][j];
      // ... Addr calculation + 123 loads continue
      temp1 = _v3;
     {double _v25; _v25 = dt * tz2; temp2 = _v25;
     {double _v435, _v434, _v283, _v433, _v436;
      _v283 = (- _v25);  _v433 = _v3 * _v432;
      _v434 = (- _v433);
      _v435 = _v434 + (_v283 * _v431);
      _v436 = _v435 - (dz1 * _v3);
      lhsZ[0][0][0][k][i][j] = _v436;
     {/* ... 74 stores */}}}
}}}
\end{lstlisting}
\vspace{-6pt}
\end{figure}

\section{Code Generation}\label{sec:codegen}

To implement the extracted solutions from e-graphs in the input code, we introduce temporary variables. During the generation step, ACC Saturator leverages a novel technique called {\it bulk load} that prioritizes high memory pressure by reordering computations.

\subsection{Temporary-Variable Insertion}
\mybox{\raisebox{-0.5em}[-0.0em][-0.0em]{\color{-.}\bfseries 3}} at Figure~\ref{fig:overview} depicts the extracted expression in a directed graph. For each selected e-node, ACC Saturator generates a temporary variable to store the computational result, placing it immediately before the corresponding use. In cases where multiple statements reference an e-node, we select the innermost scope to declare a variable for those statements. Every assignment modifies the right-hand expression to include the variable of the corresponding e-node.

The compilers of directive-based code can optimize the redundant use of registers. Our code-generation style reduces duplicate computation and leverages optimal instructions, such as FMA, while preserving ILP.

\subsection{Bulk Load}
As GPUs suffer high memory-access latency, reducing only memory accesses or computation may not lead to improved performance. ACC Saturator follows a different approach by reordering statements to increase memory pressure first and then minimizing memory operations for the remaining execution.

\begin{table}[b]
\center
\caption{NAS Parallel Benchmarks~\cite{npb}}
\label{tab:npb}
\setlength\tabcolsep{5pt}
\center
\begin{tabular}{c|cccrr}
 \multicolumn{1}{c|}{\multirow{2}{*}{Name}} &  \multicolumn{1}{c}{\multirow{2}{*}{Compute}} &  \multicolumn{1}{c}{\multirow{2}{*}{Access}} & \multicolumn{1}{c}{Num.}  & \multicolumn{2}{c}{Original Time}\\
 & & & \multicolumn{1}{c}{Kernels}  & \multicolumn{1}{l}{\footnotesize ~~NVHPC} & \multicolumn{1}{c}{\footnotesize GCC}\\\hline
\textbf{BT} & CFD & Halo (3D) & 46 & 14.85s & 28.04s \\
\textbf{CG} & Eigenvalue & Irregular & 16 & 1.27s & 26.17s \\
\textbf{EP} & Random Num & Parallel & 4 & 2.65s & 3.35s \\
\textbf{FT} & FFT & All-to-All & 12 & 3.06s & 3.10s \\
\textbf{LU} & CFD & Halo (3D) & 59 & 15.36s & 24.86s \\
\textbf{MG} & Poisson Eq & Long \& Short & 16 & 0.79s & 0.79s \\
\textbf{SP} & CFD & Halo (3D) & 65 & 10.00s & 12.00s \\
\end{tabular}
\end{table}

We address the high memory-access latency on GPUs by utilizing our
proposed technique, bulk load. This technique relocates every memory
load to the first place where its dependencies are resolved. When
multiple loads share one location, we sort these based on their static
indices. We prioritize increasing memory pressure at the beginning of
the execution and then avoiding memory operations for the rest of the
execution. To illustrate the effectiveness of this technique, we
compare the performance of a time\-/consuming kernel in the OpenACC
version of NAS Parallel Benchmarks' BT (NPB-BT) before and after
optimization by ACC Saturator. Listings~\ref{lst:zsolve}
and~\ref{lst:zsolvegen} show the original and optimized code,
respectively. Despite featuring the same directives and code
structure, the optimized code performs all the loads before the first
assignment (\texttt{temp1}), and each subsequent store refers to local
variables, leading to minimum number of operations while utilizing FMA.

\begin{table*}[t]
\vspace{-8pt}
\center
\caption{\label{tab:spec}The SPEC ACCEL benchmark suite~\cite{spec}}
\setlength\tabcolsep{5pt}
\setlength{\extrarowheight}{1pt}
\center
\begin{tabular}{c|cccllrrrrrr}
 \multicolumn{1}{c|}{\multirow{2}{*}{Name}} &  \multicolumn{1}{c}{\multirow{2}{*}{Compute}} &  \multicolumn{1}{c}{\multirow{2}{*}{Access}} & \multicolumn{1}{c}{Num.} &  \multicolumn{2}{c}{\multirow{2}{*}{Size}} & \multicolumn{2}{c}{Original Time (ACC)} & \multicolumn{3}{c}{Original Time (OMP)}\\
 & & & \multicolumn{1}{c}{kernels}  & &  & \multicolumn{1}{l}{\footnotesize \hspace{1.4em}NVHPC} & \multicolumn{1}{c}{\footnotesize  \hspace{1.4em}GCC} & \multicolumn{1}{l}{\footnotesize \hspace{0.7em}NVHPC} & \multicolumn{1}{c}{\footnotesize GCC} & \multicolumn{1}{c}{\footnotesize Clang}\\\hline
\textbf{ostencil} & Jacobi & Halo (3D) & 1 & \multicolumn{2}{c}{Ref} & 3.87s & 10.28s & 7.75s & 107.54s & 34.60s \\
\textbf{olbm} & CFD & Halo (3D) & 3 & \multicolumn{2}{c}{Ref} & 7.11s & 13.32s & 7.11s & 13.47s & 5.91s \\
\textbf{omriq} & MRI & Structure-of-arrays & 2 & \multicolumn{2}{c}{Ref} & 16.02s  & 16.18s & 5.99s & 18.54s & 11.87s \\
\textbf{ep} & Random Num & Parallel & 5 &  Ref / Test & (CLASS D / W) & 45.33s & 69.91s & 62.42s & 90.35s & 71.32s \\
\textbf{cg} & Eigenvalue & Irregular & 16 & Ref & (> CLASS C) & 4.28s & 662.58s & 5.06s & 19.03s & 18.42s \\
\textbf{csp} & CFD & Halo (3D) & 68 & Ref / Test & (CLASS C / S) & 7.71s & 27.26s & 111.79s & 589.87s & 105.75s \\
\textbf{bt} & CFD & Halo (3D) & 50 & Ref / Test & (CLASS B / W) & 3.24s & 130.43s & 555.44s & 60.45s & 562.83s \\
\end{tabular}
\end{table*}

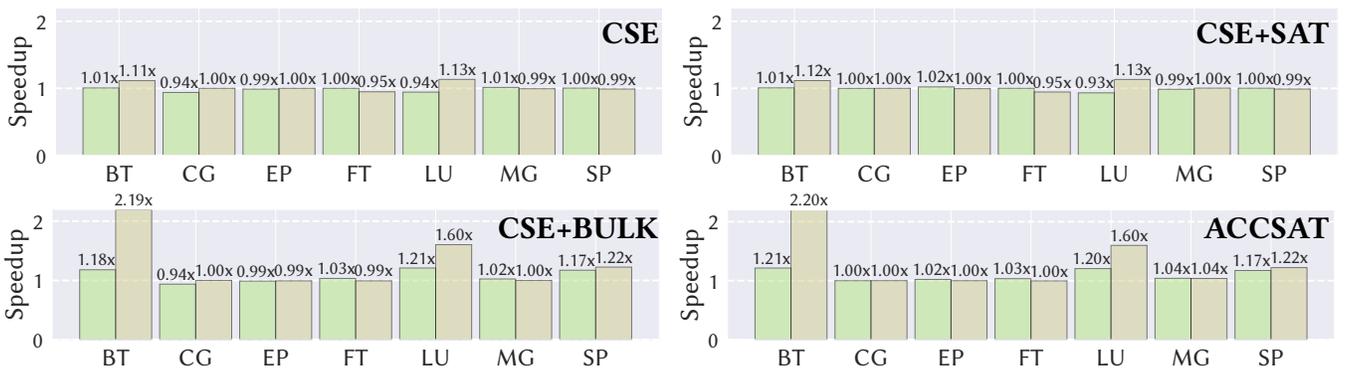
\begin{figure*}[b]
  \center
  \vspace*{0.10cm}
  \noindent\makebox[0pt]{
  \includegraphics[width=0.5\textwidth]{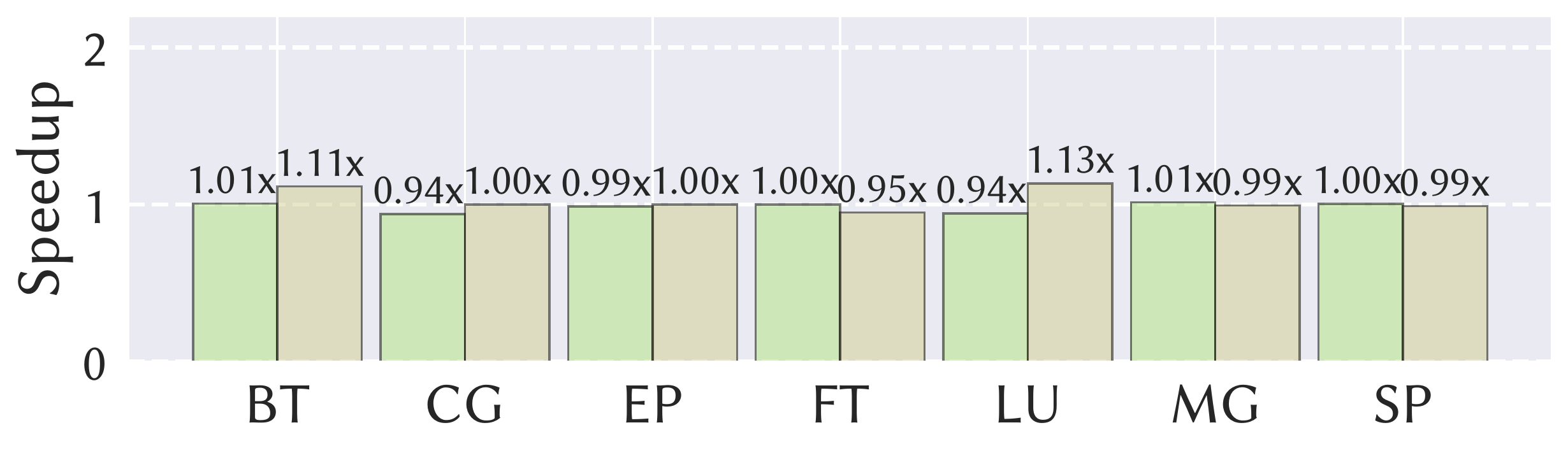}
  \includegraphics[width=0.5\textwidth]{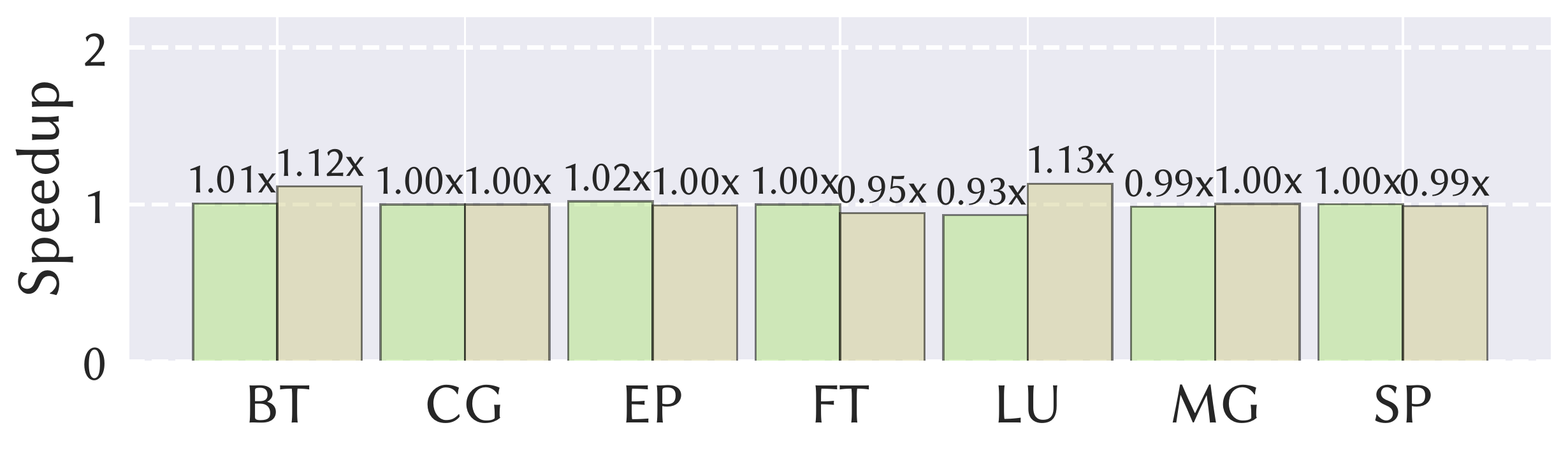}
  }
  \vspace{-0.15cm}\\
  \center
  \noindent\makebox[0pt]{
  \includegraphics[width=0.5\textwidth]{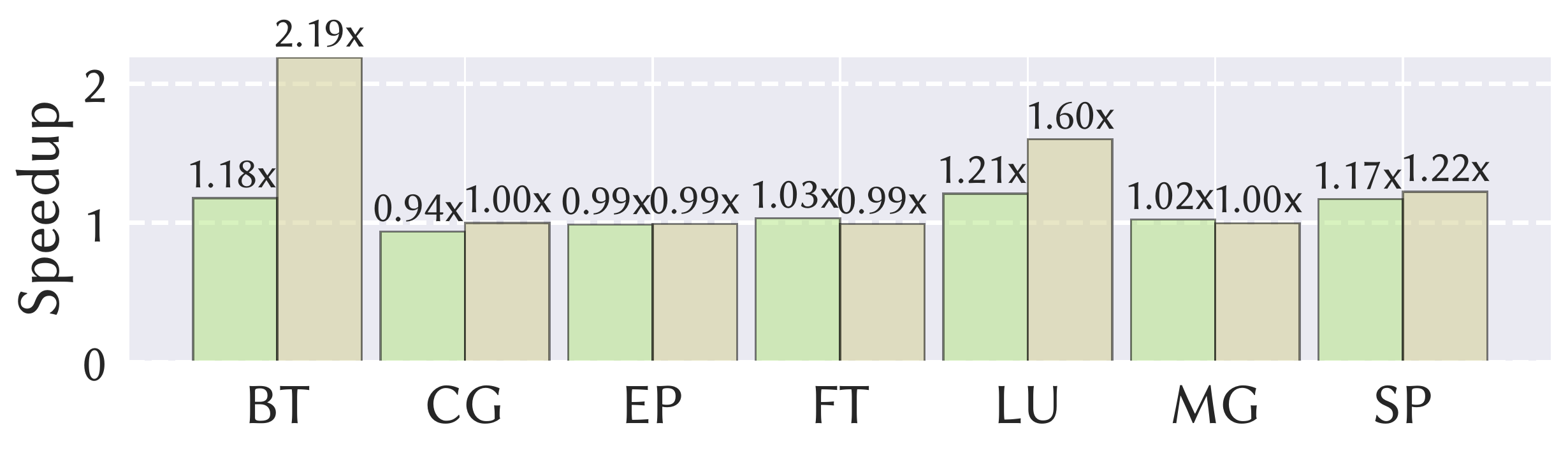}
  \includegraphics[width=0.5\textwidth]{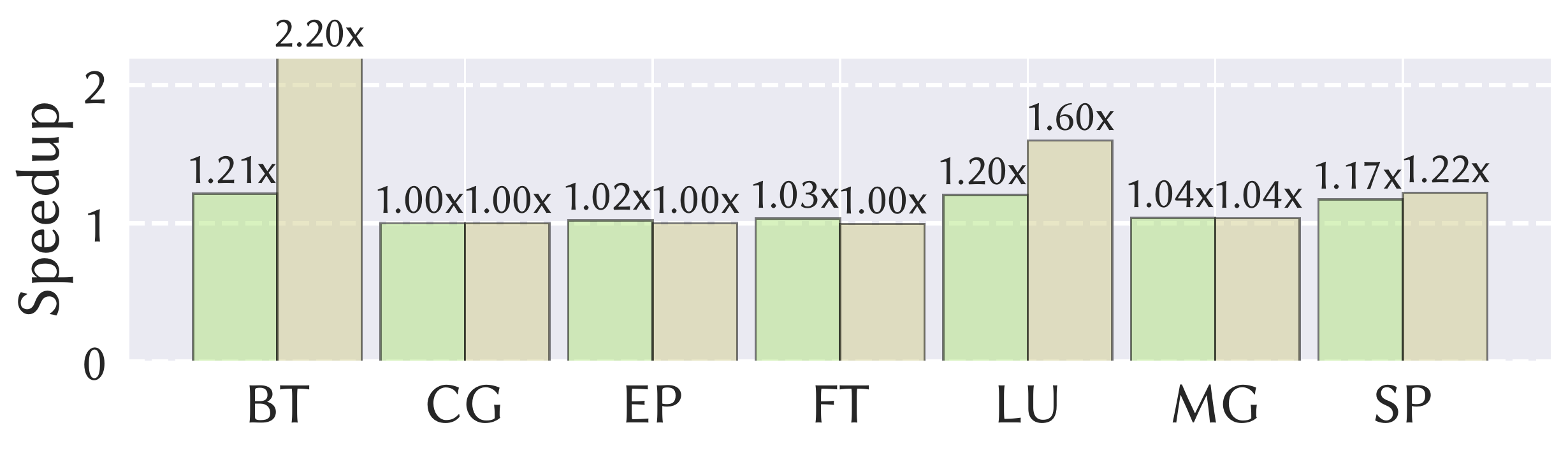}
  }
\begin{tikzpicture}[remember picture,overlay,>=stealth]
  \node[font=\bfseries] at (-0.8,4.8) {\large CSE};
  \node[font=\bfseries] at (-1.5,2) {\large CSE+BULK};
  \node[font=\bfseries] at (7.8,4.8) {\large CSE+SAT};
  \node[font=\bfseries] at (7.85,2) {\large ACCSAT};
\end{tikzpicture}
  \vspace*{-0.35cm}
  \caption{NPB's speedup results on NVIDIA A100-PCIE-40GB for each variation compared to original. {\color{colornvhpc}$\squadfill$}~NVHPC, {\color{colorgcc}$\squadfill$}~GCC.}
  \label{fig:npb}
  \vspace*{0.0cm}
\end{figure*}

\section{Experimental Methodology}\label{sec:method}

We implement ACC Saturator in Racket~\cite{racket} using XcodeML~\cite{xcodeml} to parse and generate OpenACC/OpenMP source codes in C.
We use the egg library~\cite{egg} to perform equality saturation.
ACC Saturator integrates with NVHPC~\cite{nvhpc} and GCC~\cite{gcc} for OpenACC/OpenMP compilation, and with Clang~\cite{clang} for OpenMP compilation.
For evaluation, we use NVHPC 22.9 with options \ctext{\texttt{"-O3 -gpu=fastmath -Msafeptr -(acc|mp)=gpu"}}, GCC 12.2.0 with \ctext{\texttt{"-O3 -ffast-math -f(openacc|openmp)"}} and Clang 15.0.3 with \ctext{\texttt{"-O3 -ffast-math -fopenmp"}}.
Our experiments run on an NVIDIA A100-PCIE-40GB GPU with an Intel Xeon Silver 4114 CPU.

We evaluate the kernel-execution performance of ACC Saturator on two benchmark suites:
NAS Parallel Benchmarks in OpenACC/C (NPB)~\cite{npb} and the SPEC ACCEL benchmark suite in both OpenACC/C and OpenMP/C (SPEC)~\cite{spec}.
Tables~\ref{tab:npb} and~\ref{tab:spec} provide the detail of NPB and SPEC, respectively.
To ensure adequate memory usage, we select CLASS C as the problem
sizes of all NPB benchmarks (the largest size within standard test problems),
SPEC uses the referential sizes (Ref) except for GCC's OpenACC cases
of \textbf{ep}, \textbf{sp}, and \textbf{bt}, for which we select the testing sizes (Test) due to high execution latency.
NPB's \textbf{BT}, \textbf{CG}, \textbf{EP}, and \textbf{SP} feature the same computation as SPEC's \textbf{bt}, \textbf{cg}, \textbf{ep}, and \textbf{csp}, but the implementation of NPB is based on OpenACC's parallel directive while that of SPEC's OpenACC benchmarks is on the kernels directive.
We report the best kernel performance of three executions.
To avoid producing incorrect results, we remove the user-specific parallelism in NPB's \textbf{CG} for GCC, disable the device-side reduction of GCC's OpenACC in NPB's \textbf{LU} and \textbf{MG},
and omit the degree specification of the worker parallelism from GCC's
NPB cases, since it surpasses GCC's thread limit.

On average, ACC Saturator spends 91.8 ms ($\sigma$ = 253.3; 1.4 $\sim$ 1885.0 ms) to construct SSA and generate code for each kernel.
Equality saturation is executed within a limit of 10,000 e-nodes, 10 seconds of saturation time, 10 rewriting iterations, and a 30-second extraction time limit. The results of the benchmarks show that each kernel requires an average of 0.63 sec ($\sigma$ = 3.37; 0.00 $\sim$ 31.2 sec) for equality saturation.

\section{Evaluation}\label{sec:evaluation}

\begin{table*}[t]
\vspace{0pt}
\def\arraystretch{1}
\setlength\tabcolsep{2pt}
\setlength{\extrarowheight}{1pt}
\scriptsize
\center
\caption{\label{tab:breakdown}Top-10 kernel breakdown of NPB-BT}
\center
\begin{tabular}{c|rrrrr|rrrrr|rrrrr|rrrrr|rrrrr}
  & \multicolumn{5}{c|}{\bf NVHPC (14.85 sec)} & \multicolumn{5}{c|}{\bf + CSE (14.77 sec) \ \ \ {\color{red}$\circletfill$}} & \multicolumn{5}{c|}{\bf + CSE+SAT (14.75 sec) \ \ \ $\circletfill$} & \multicolumn{5}{c|}{\bf + CSE+BULK (12.59 sec) \ \ \ {\color{coloronlybulk}$\pentagofill$}} & \multicolumn{5}{c}{\bf + ACCSAT (12.23 sec) \ \ \ {\color{colorsat}$\pentagofill$}} \\
  \% & \multicolumn{1}{c}{\faIcon{clock}} & \multicolumn{1}{c}{\faIcon{microchip}} & \multicolumn{1}{c}{\faIcon{memory}} & \multicolumn{1}{c}{\faIcon{cogs}} & \multicolumn{1}{c|}{\faIcon{chart-pie}} & \multicolumn{1}{c}{\faIcon{clock}} & \multicolumn{1}{c}{\faIcon{microchip}} & \multicolumn{1}{c}{\faIcon{memory}} & \multicolumn{1}{c}{\faIcon{cogs}} & \multicolumn{1}{c|}{\faIcon{chart-pie}} & \multicolumn{1}{c}{\faIcon{clock}} & \multicolumn{1}{c}{\faIcon{microchip}} & \multicolumn{1}{c}{\faIcon{memory}} & \multicolumn{1}{c}{\faIcon{cogs}} & \multicolumn{1}{c|}{\faIcon{chart-pie}} & \multicolumn{1}{c}{\faIcon{clock}} & \multicolumn{1}{c}{\faIcon{microchip}} & \multicolumn{1}{c}{\faIcon{memory}} & \multicolumn{1}{c}{\faIcon{cogs}} & \multicolumn{1}{c|}{\faIcon{chart-pie}} & \multicolumn{1}{c}{\faIcon{clock}} & \multicolumn{1}{c}{\faIcon{microchip}} & \multicolumn{1}{c}{\faIcon{memory}} & \multicolumn{1}{c}{\faIcon{cogs}} & \multicolumn{1}{c}{\faIcon{chart-pie}}  \\
\hline
13.6\% & 10.08 & 78.0 & 34.50\% & 152 & \bf{0.19} & 10.05 & 100\% & +0.68\% & \bf{-2} & +0.00 & 10.04 & \bf{99\%} & +0.22\% & +10 & +0.00 & \bf{6.40} & 106\% & \bf{+38.00\%} & +103 & -0.06 & 6.57 & 104\% & +37.26\% & +103 & -0.06\\
13.6\% & 10.07 & 78.0 & 34.58\% & 152 & \bf{0.19} & 10.06 & 100\% & -0.74\% & \bf{-2} & +0.00 & 10.05 & \bf{99\%} & +0.12\% & +10 & +0.00 & \bf{6.40} & 106\% & \bf{+38.02\%} & +103 & -0.06 & 6.56 & 104\% & +37.26\% & +103 & -0.06\\
13.5\% & 9.98 & 78.0 & 35.86\% & 152 & \bf{0.12} & 9.96 & 100\% & -0.09\% & \bf{-2} & +0.00 & 9.97 & \bf{99\%} & -0.07\% & +10 & +0.00 & \bf{6.69} & 106\% & \bf{+32.68\%} & +103 & +0.00 & 6.72 & 104\% & +32.33\% & +103 & +0.00\\
8.8\% & 6.48 & 116.3 & 54.06\% & \bf{176} & \bf{0.12} & 6.42 & 100\% & +0.02\% & +0 & +0.00 & 6.43 & 100\% & +0.01\% & +2 & +0.00 & 6.85 & 108\% & -2.02\% & +78 & +0.00 & \bf{6.28} & \bf{99\%} & \bf{+2.02\%} & +79 & +0.00\\
8.1\% & \bf{6.01} & 116.3 & \bf{57.11\%} & \bf{176} & \bf{0.12} & 6.05 & 100\% & -0.14\% & +2 & +0.00 & 6.06 & 100\% & -0.40\% & +2 & +0.00 & 7.14 & 108\% & -6.69\% & +78 & +0.00 & 6.39 & \bf{99\%} & -0.01\% & +79 & +0.00\\
8.0\% & \bf{5.92} & 116.3 & 57.32\% & \bf{176} & \bf{0.12} & 5.96 & 100\% & -0.25\% & +2 & +0.00 & 5.98 & 100\% & -0.57\% & +2 & +0.00 & 7.08 & 108\% & -5.92\% & +78 & +0.00 & 6.33 & \bf{99\%} & \bf{+0.11\%} & +79 & +0.00\\
5.1\% & 0.023 & 0.1 & 39.62\% & \bf{47} & \bf{0.62} & 0.024 & 103\% & -1.04\% & +0 & +0.00 & 0.024 & 103\% & -0.03\% & +0 & +0.00 & 0.018 & \bf{99\%} & +5.53\% & +39 & -0.31 & \bf{0.018} & 99\% & \bf{+6.86\%} & +39 & -0.31\\
5.0\% & 3.66 & 35.6 & 52.23\% & 96 & \bf{0.25} & 3.63 & \bf{97\%} & +1.37\% & -2 & +0.00 & 3.61 & 99\% & +1.28\% & \bf{-4} & +0.00 & \bf{3.38} & 97\% & +2.74\% & +0 & +0.00 & 3.38 & 99\% & \bf{+3.43\%} & +12 & +0.00\\
4.6\% & 3.39 & 34.9 & 52.50\% & \bf{90} & \bf{0.25} & 3.06 & \bf{96\%} & -0.70\% & +23 & +0.00 & 3.07 & 98\% & -0.77\% & +27 & +0.00 & 3.03 & 96\% & \bf{+0.61\%} & +23 & +0.00 & \bf{3.00} & 98\% & +0.21\% & +27 & +0.00\\
4.0\% & 0.018 & 0.1 & 44.22\% & \bf{47} & \bf{0.62} & 0.019 & 103\% & +0.54\% & +0 & +0.00 & 0.018 & 103\% & +0.01\% & +0 & +0.00 & 0.016 & \bf{99\%} & +2.83\% & +39 & -0.31 & \bf{0.016} & 99\% & \bf{+4.37\%} & +39 & -0.31\\
\end{tabular}

\vspace*{1em}
\begin{tabular}{c|rrrrr|rrrrr|rrrrr|rrrrr|rrrrr}
  & \multicolumn{5}{c|}{\bf GCC (28.04 sec)} & \multicolumn{5}{c|}{\bf + CSE (25.16 sec) \ \ \ {\color{red}$\circletfill$}} & \multicolumn{5}{c|}{\bf + CSE+SAT (25.13 sec) \ \ \ $\circletfill$} & \multicolumn{5}{c|}{\bf + CSE+BULK (12.79 sec) \ \ \ {\color{coloronlybulk}$\pentagofill$}} & \multicolumn{5}{c}{\bf + ACCSAT (12.74 sec) \ \ \ {\color{colorsat}$\pentagofill$}} \\
  \% & \multicolumn{1}{c}{\faIcon{clock}} & \multicolumn{1}{c}{\faIcon{microchip}} & \multicolumn{1}{c}{\faIcon{memory}} & \multicolumn{1}{c}{\faIcon{cogs}} & \multicolumn{1}{c|}{\faIcon{chart-pie}} & \multicolumn{1}{c}{\faIcon{clock}} & \multicolumn{1}{c}{\faIcon{microchip}} & \multicolumn{1}{c}{\faIcon{memory}} & \multicolumn{1}{c}{\faIcon{cogs}} & \multicolumn{1}{c|}{\faIcon{chart-pie}} & \multicolumn{1}{c}{\faIcon{clock}} & \multicolumn{1}{c}{\faIcon{microchip}} & \multicolumn{1}{c}{\faIcon{memory}} & \multicolumn{1}{c}{\faIcon{cogs}} & \multicolumn{1}{c|}{\faIcon{chart-pie}} & \multicolumn{1}{c}{\faIcon{clock}} & \multicolumn{1}{c}{\faIcon{microchip}} & \multicolumn{1}{c}{\faIcon{memory}} & \multicolumn{1}{c}{\faIcon{cogs}} & \multicolumn{1}{c|}{\faIcon{chart-pie}} & \multicolumn{1}{c}{\faIcon{clock}} & \multicolumn{1}{c}{\faIcon{microchip}} & \multicolumn{1}{c}{\faIcon{memory}} & \multicolumn{1}{c}{\faIcon{cogs}} & \multicolumn{1}{c}{\faIcon{chart-pie}}  \\
\hline
20.1\% & 28.01 & 226.2 & 21.66\% & \bf{134} & \bf{0.19} & 24.29 & 89\% & +2.40\% & +0 & +0.00 & 24.26 & 89\% & +2.43\% & +0 & +0.00 & 5.53 & \bf{50\%} & \bf{+41.55\%} & +121 & -0.06 & \bf{5.51} & 50\% & +41.09\% & +121 & -0.06\\
20.1\% & 28.06 & 226.2 & 19.81\% & \bf{130} & \bf{0.19} & 23.80 & 89\% & +2.98\% & +0 & +0.00 & 23.67 & 89\% & +3.00\% & +0 & +0.00 & 5.59 & \bf{50\%} & +43.07\% & +125 & -0.06 & \bf{5.58} & 50\% & \bf{+43.41\%} & +125 & -0.06\\
20.0\% & 27.93 & 226.2 & 20.95\% & \bf{134} & \bf{0.19} & 24.20 & 89\% & +2.44\% & +2 & +0.00 & 24.16 & 89\% & +2.42\% & +0 & +0.00 & \bf{5.46} & 50\% & \bf{+42.48\%} & +121 & -0.06 & 5.50 & \bf{50\%} & +42.16\% & +121 & -0.06\\
5.4\% & 7.54 & 97.5 & 56.38\% & \bf{98} & \bf{0.25} & 7.54 & 100\% & +0.06\% & +0 & +0.00 & 7.55 & \bf{99\%} & +0.49\% & +0 & +0.00 & \bf{6.74} & 101\% & \bf{+12.70\%} & +157 & -0.12 & 6.78 & 100\% & +12.17\% & +157 & -0.12\\
5.4\% & 7.53 & 97.5 & 56.42\% & \bf{98} & \bf{0.25} & 7.53 & 100\% & -0.04\% & +0 & +0.00 & 7.54 & \bf{99\%} & -0.02\% & +0 & +0.00 & \bf{6.74} & 100\% & \bf{+12.35\%} & +157 & -0.12 & 6.77 & 100\% & +12.29\% & +157 & -0.12\\
5.4\% & 7.54 & 97.5 & 56.77\% & 98 & 0.25 & 7.54 & 100\% & -0.76\% & +0 & +0.00 & 7.54 & \bf{99\%} & -0.47\% & +0 & +0.00 & \bf{6.76} & 100\% & +11.93\% & +157 & -0.12 & 6.82 & 100\% & \bf{+12.10\%} & +157 & -0.12\\
3.9\% & 0.034 & 0.3 & \bf{43.50\%} & \bf{56} & \bf{0.56} & 0.034 & 100\% & -0.35\% & +0 & +0.00 & 0.034 & 100\% & -0.44\% & +0 & +0.00 & 0.025 & \bf{92\%} & -9.16\% & +48 & -0.31 & \bf{0.025} & 92\% & -8.73\% & +48 & -0.31\\
3.4\% & 4.71 & 37.8 & 48.48\% & \bf{96} & \bf{0.31} & 3.58 & 87\% & -1.79\% & +0 & +0.00 & 3.60 & 86\% & -3.22\% & +0 & +0.00 & 3.58 & 87\% & -2.30\% & +0 & +0.00 & \bf{3.43} & \bf{85\%} & \bf{+0.24\%} & +26 & -0.06\\
3.3\% & 4.64 & 39.9 & 41.44\% & \bf{96} & \bf{0.31} & 4.03 & 87\% & \bf{+8.93\%} & +8 & -0.06 & 4.03 & 87\% & +8.43\% & +10 & -0.06 & \bf{3.88} & 87\% & +8.11\% & +12 & -0.06 & 3.88 & \bf{87\%} & +7.09\% & +10 & -0.06\\
3.2\% & 4.50 & 39.1 & \bf{49.96\%} & \bf{96} & \bf{0.31} & 3.56 & 88\% & -3.55\% & +0 & +0.00 & 3.54 & \bf{86\%} & -4.09\% & +0 & +0.00 & 3.58 & 89\% & -4.06\% & +0 & +0.00 & \bf{3.43} & 87\% & -1.04\% & +14 & -0.06\\
\end{tabular}
\center
{\faIcon{clock} indicates the average execution time per launch (ms), \faIcon{microchip} the number of executed instructions (x$10^6$), \faIcon{memory} the memory utilization, \faIcon{cogs} the number of registers per thread, and \faIcon{chart-pie} the SM occupancy. In the columns of optimization, the last four use relative numbers comparing to the original. The most reduced/utilized numbers are shown bold.}
\end{table*}

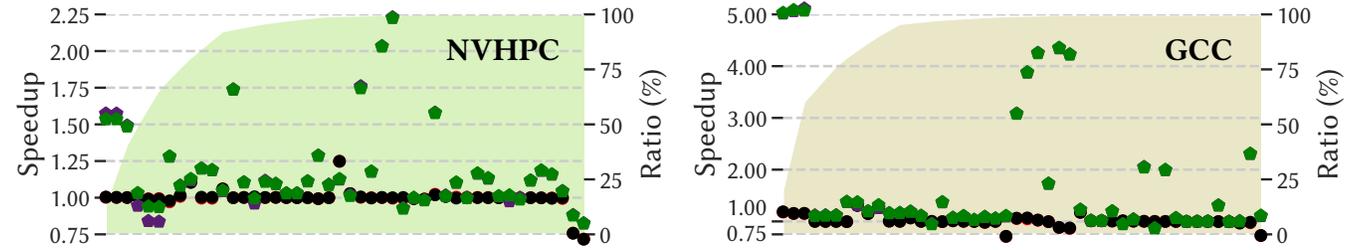
\begin{figure*}[t]
  \centering
  \noindent\makebox[\textwidth]{
    \hspace*{0.1em}
    \includegraphics[height=9.9em]{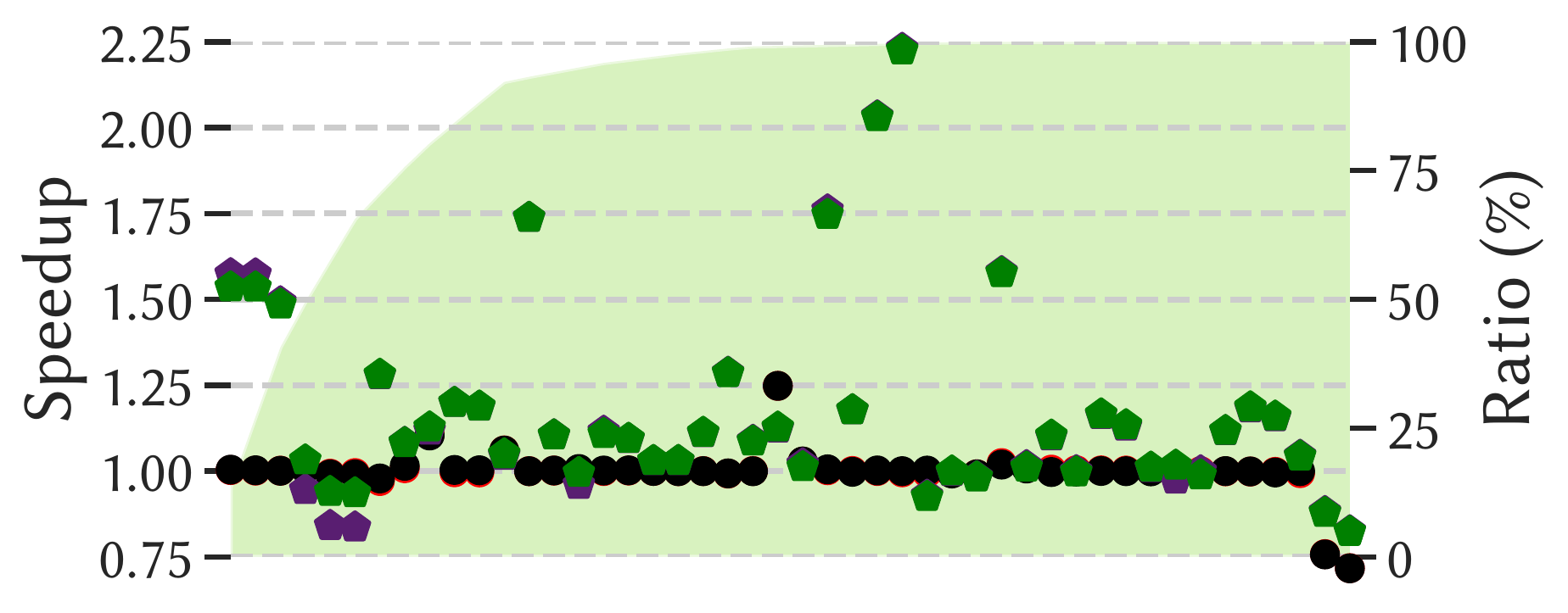}
    \hspace*{-0.2em}
    \includegraphics[height=9.9em]{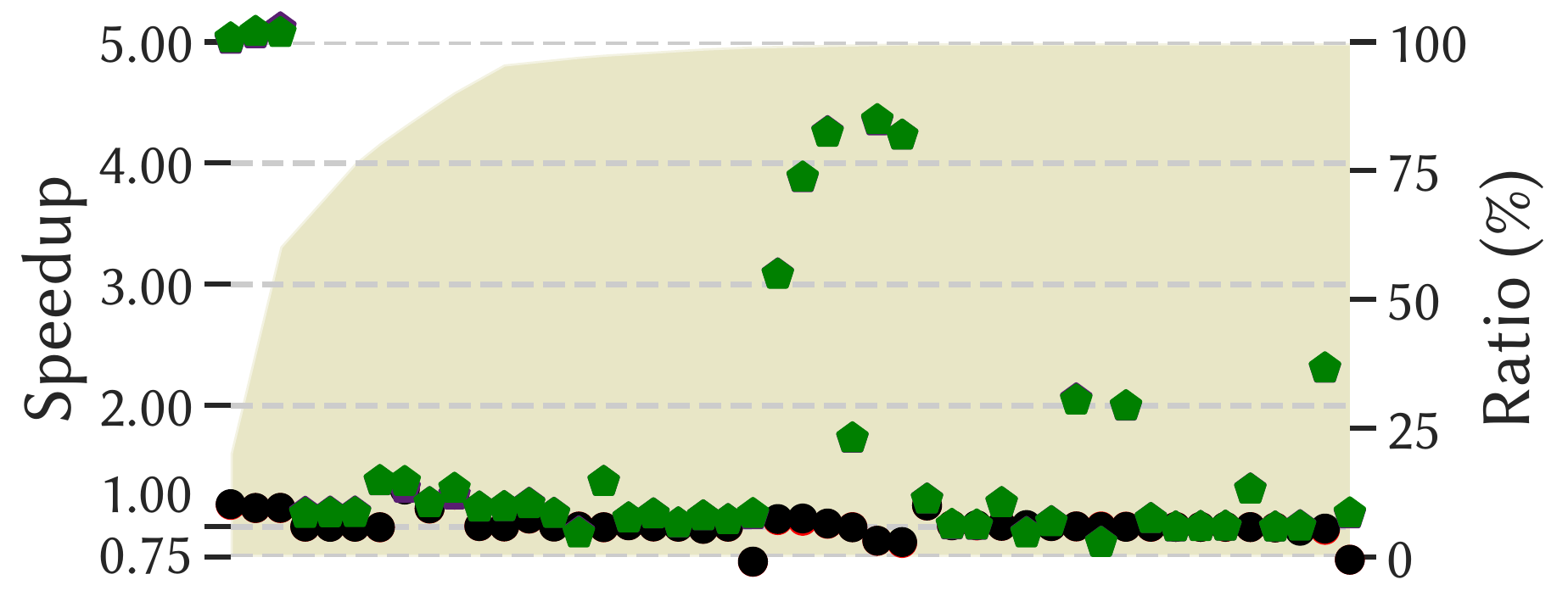}
  }
\begin{tikzpicture}[remember picture,overlay,>=stealth]
  \node[font=\bfseries] at (-2.25,3.2) {\large NVHPC};
  \node[font=\bfseries] at (7,3.2) {\large GCC};
\end{tikzpicture}
  \vspace*{-0.8cm}
  \caption{Breakdown of NPB-BT; The background color depicts the cumulative ratio of the execution time along the speedup points for each kernel. {\color{red}$\circletfill$}~\textbf{CSE}, $\circletfill$~\textbf{CSE+SAT}, {\color{coloronlybulk}$\pentagofill$}~\textbf{CSE+BULK}, {\color{colorsat}$\pentagofill$}~\textbf{ACCSAT}~(CSE+SAT+BULK).}
  \label{fig:breakdown}
\end{figure*}

\begin{figure*}[t]
  \vspace*{0.0cm}
  \centering
  \noindent\makebox[\textwidth]{
  \includegraphics[width=0.5\textwidth]{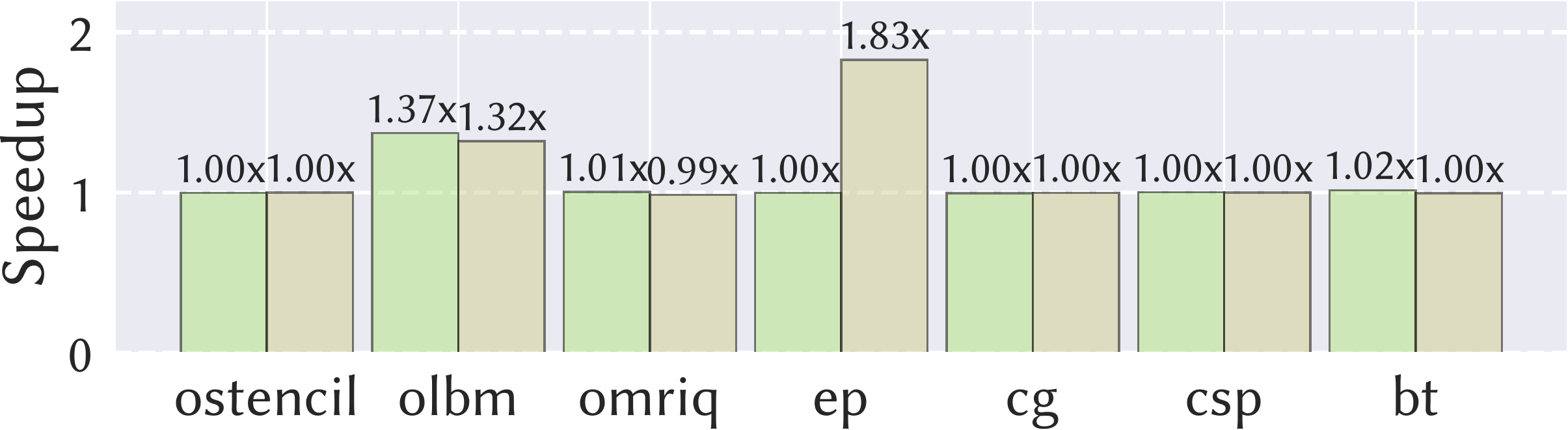}
  \hspace*{-0.8em}
  \includegraphics[width=0.5\textwidth]{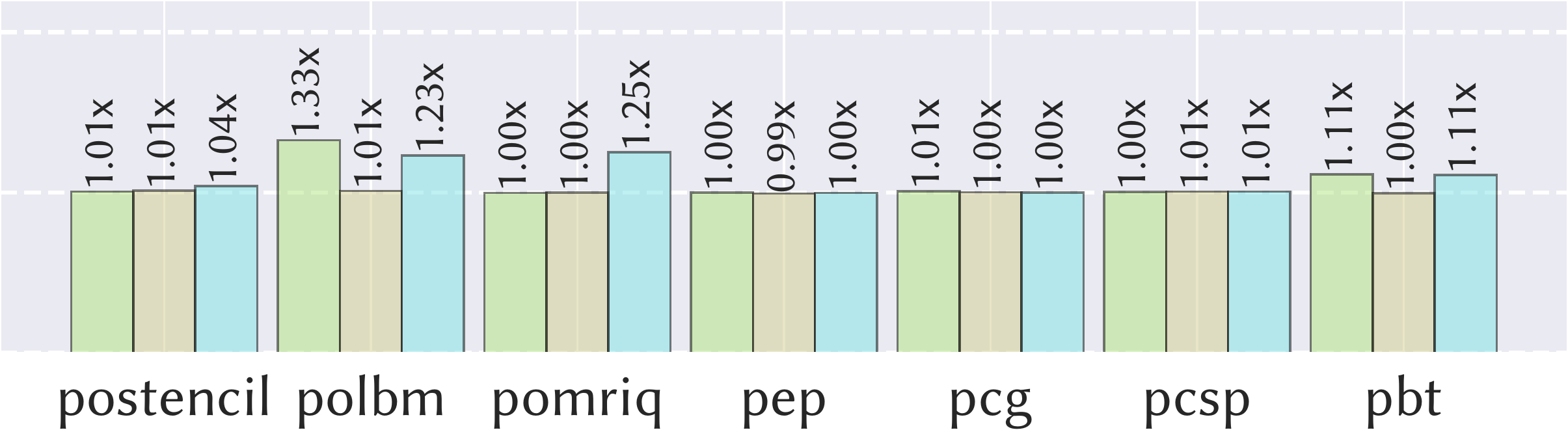}
  }
  \noindent\makebox[\textwidth]{
  \includegraphics[width=0.5\textwidth]{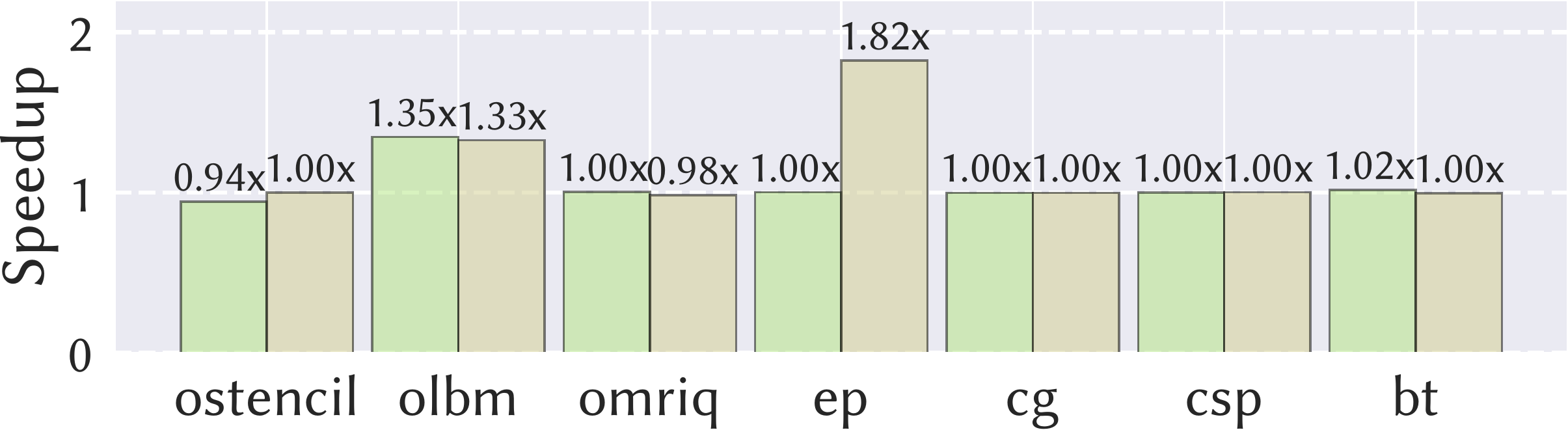}
  \hspace*{-0.8em}
  \includegraphics[width=0.5\textwidth]{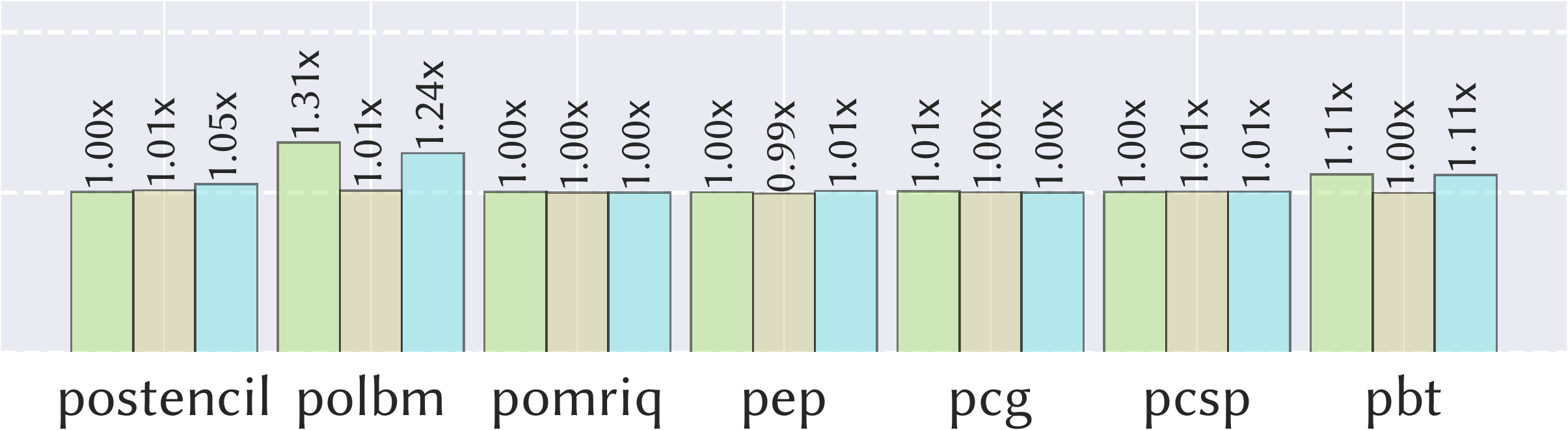}
  }
  \noindent\makebox[\textwidth]{
  \includegraphics[width=0.509\textwidth]{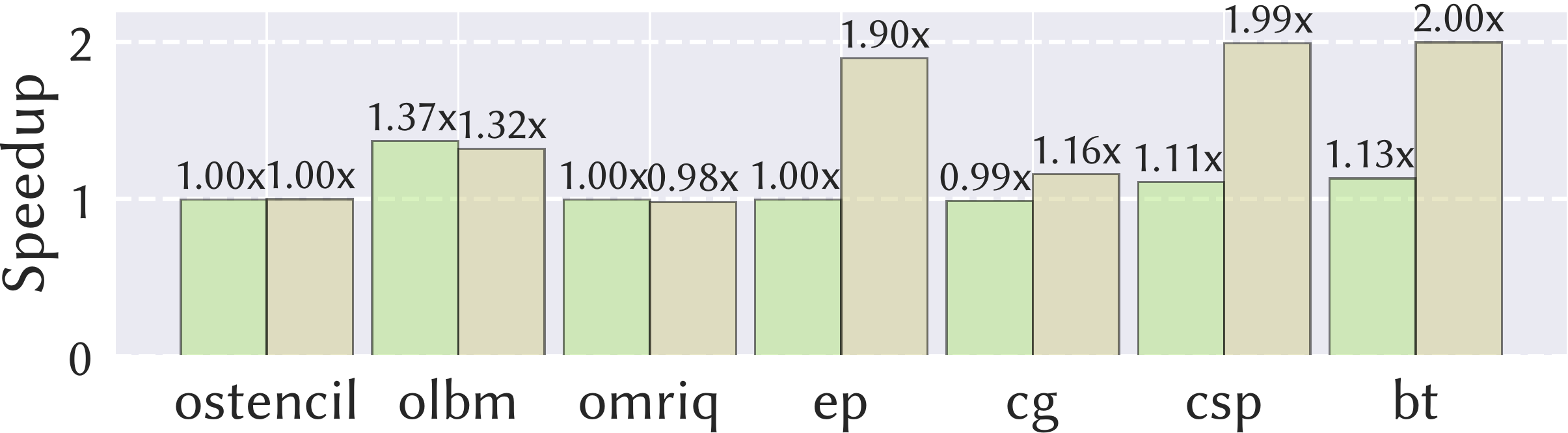}
  \hspace*{-0.8em}
  \includegraphics[width=0.5\textwidth]{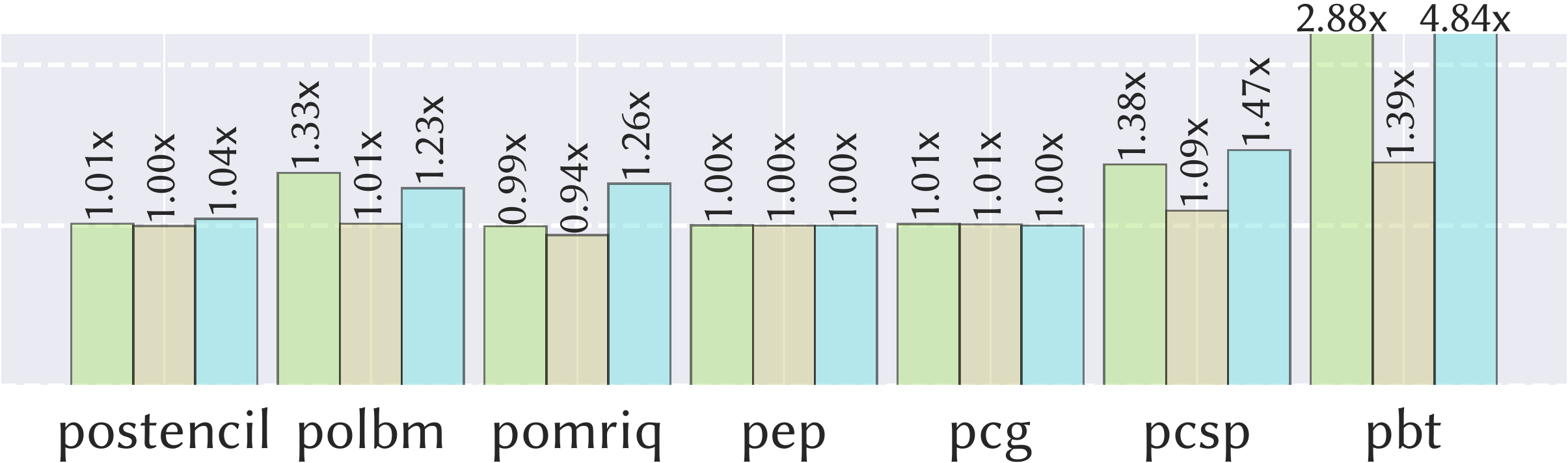}
  }
  \noindent\makebox[\textwidth]{
  \includegraphics[width=0.508\textwidth]{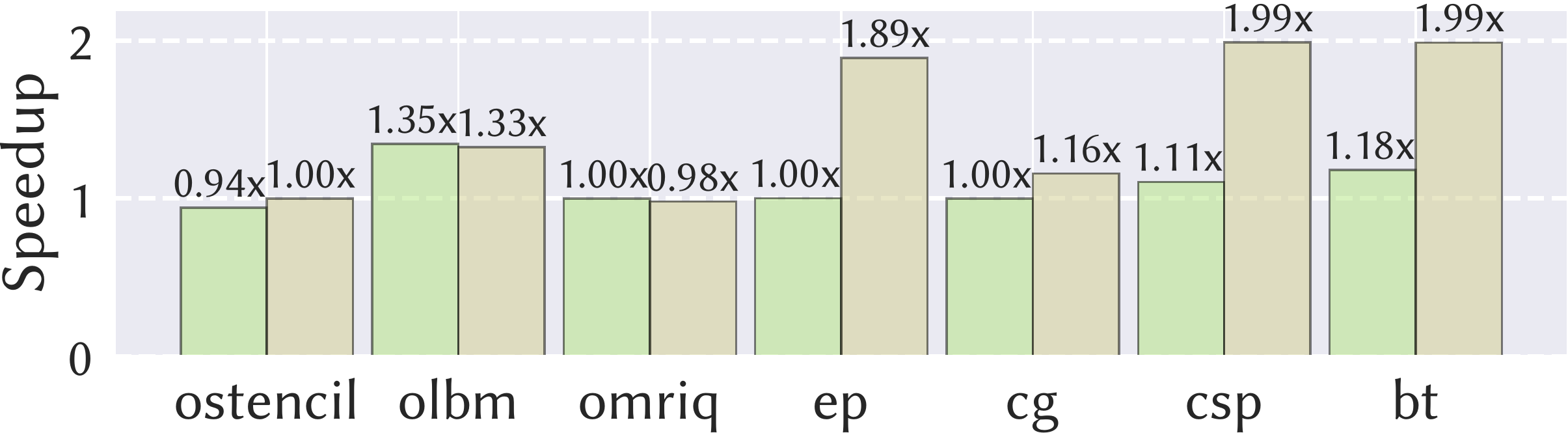}
  \hspace*{-0.8em}
  \includegraphics[width=0.5\textwidth]{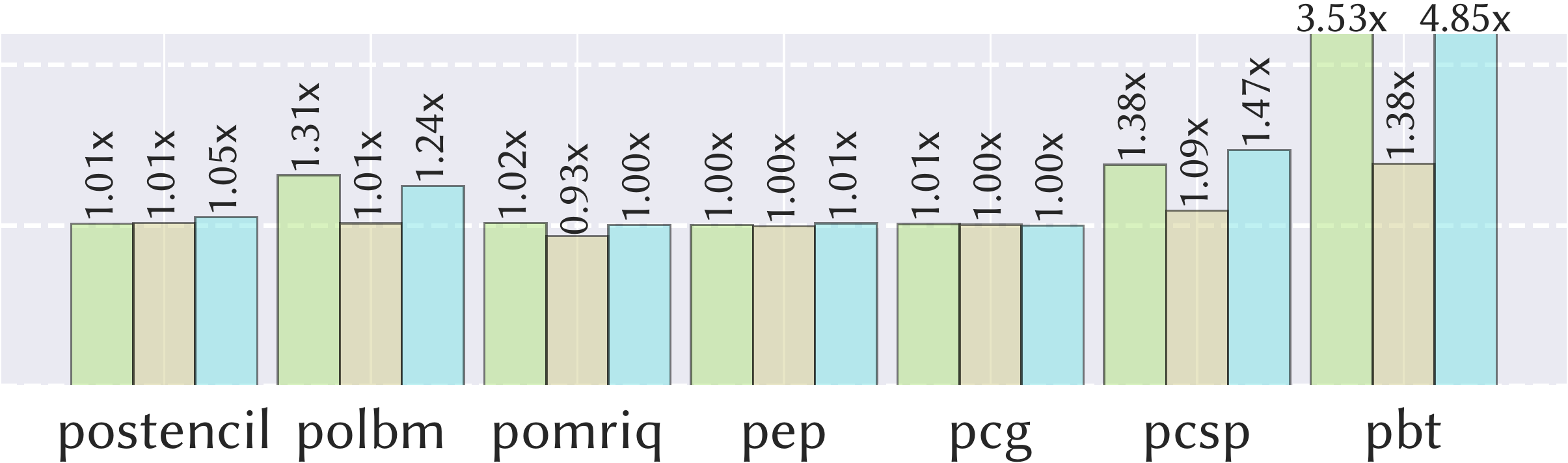}
  }
\begin{tikzpicture}[remember picture,overlay,>=stealth]
  \node[font=\bfseries] at (-7.61,10.35) {\large CSE};
  \node[font=\bfseries] at (-7.12,7.9) {\large CSE+SAT};
  \node[font=\bfseries] at (-6.94,5.25) {\large CSE+BULK};
  \node[font=\bfseries] at (-7.2,2.6) {\large ACCSAT};
\end{tikzpicture}
  \vspace*{-0.7cm}
  \caption{Speedup results of the SPEC ACCEL benchmark suite on NVIDIA A100-PCIE-40GB. {\color{colornvhpc}$\squadfill$}~NVHPC, {\color{colorgcc}$\squadfill$}~GCC, {\color{colorclang}$\squadfill$}~Clang.}
  \label{fig:spec}
  \vspace*{0.0cm}
\end{figure*}
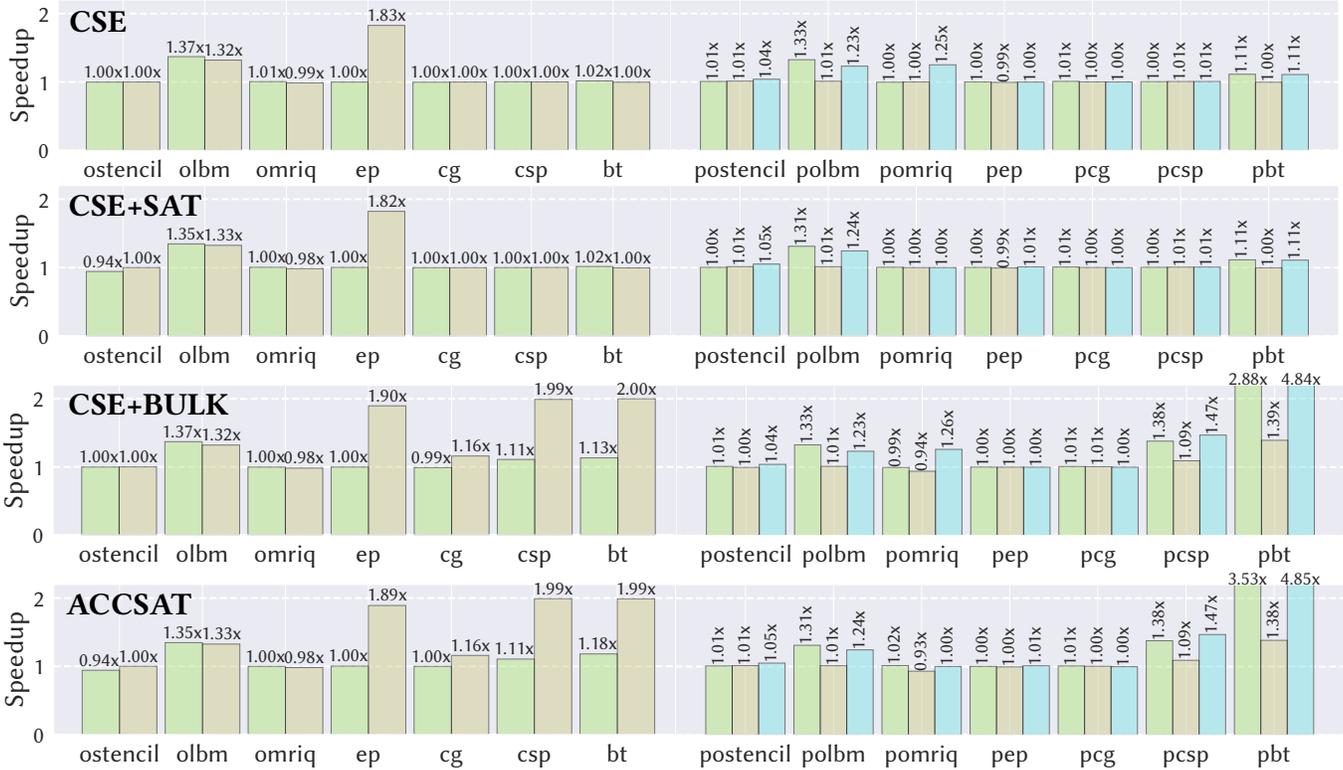

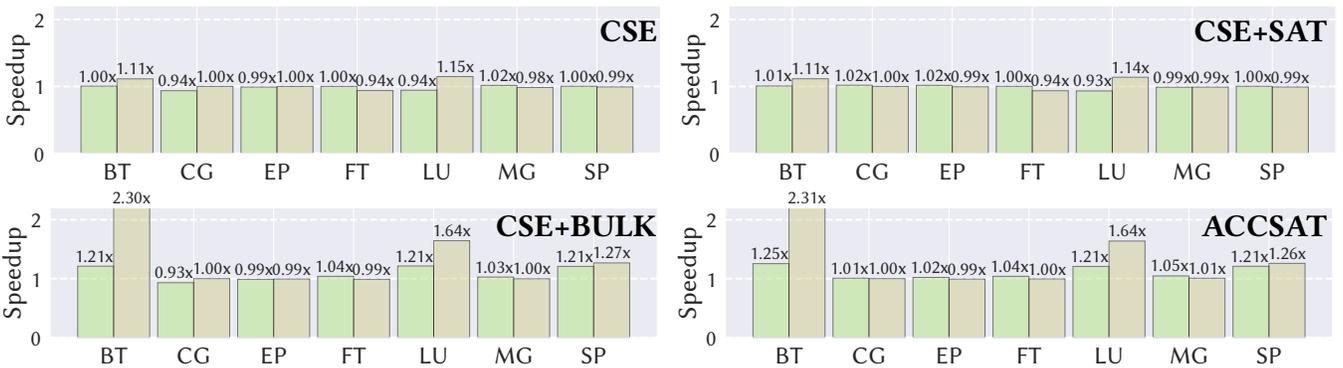
\begin{figure*}[b]
  \center
  \vspace*{0.0cm}
  \noindent\makebox[0pt]{
  \includegraphics[width=0.5\textwidth]{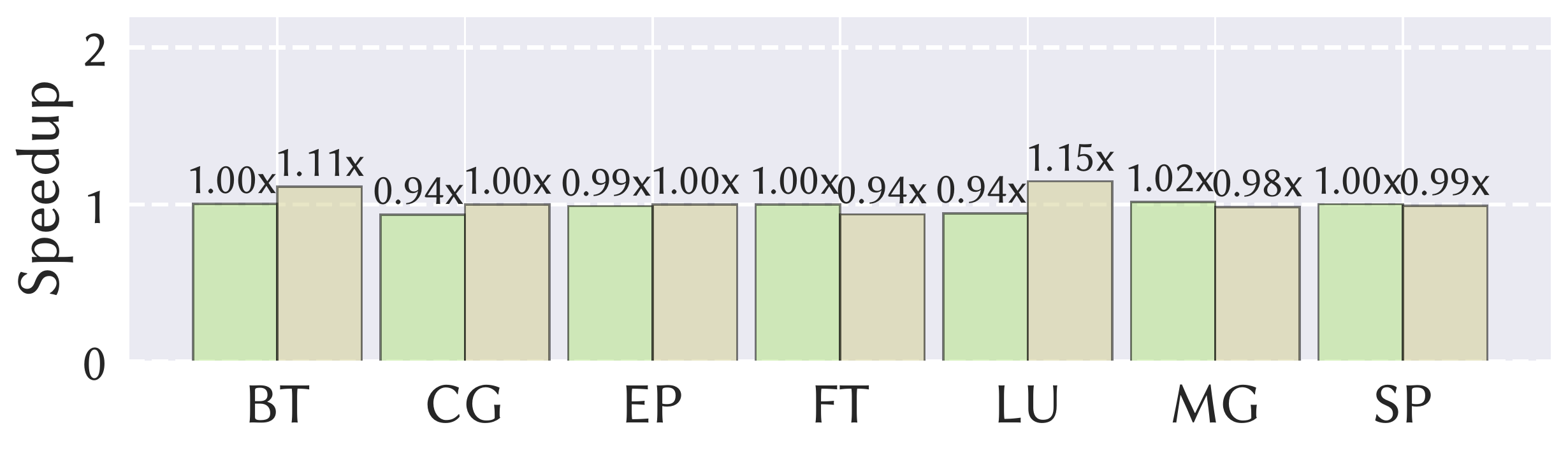}
  \includegraphics[width=0.5\textwidth]{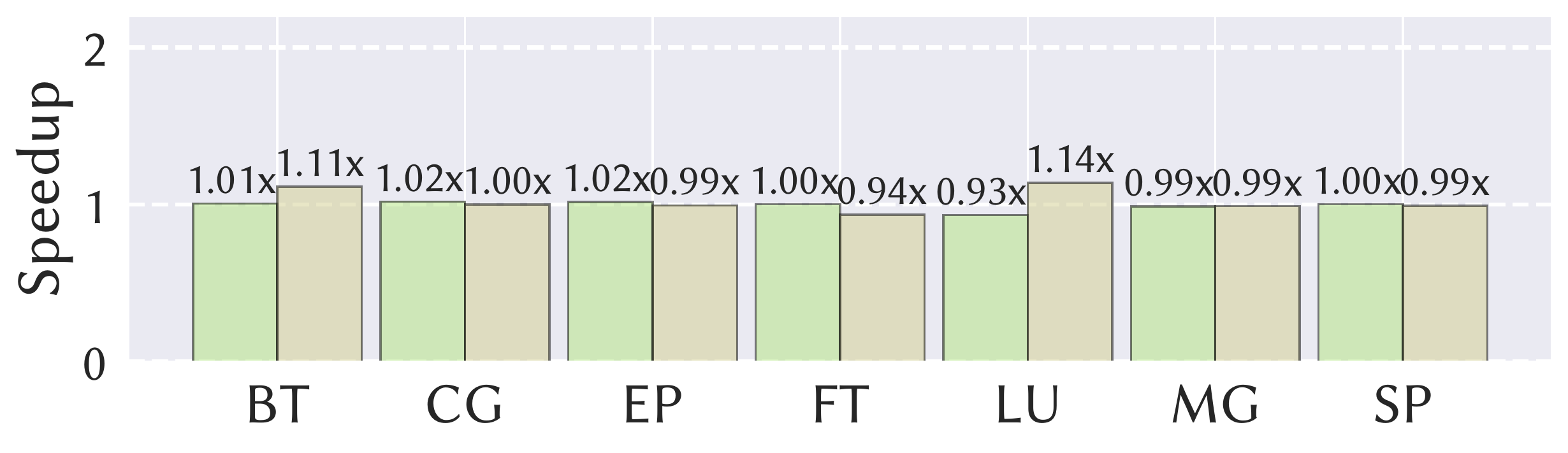}
  }
  \vspace{-0.15cm}\\
  \center
  \noindent\makebox[0pt]{
  \includegraphics[width=0.5\textwidth]{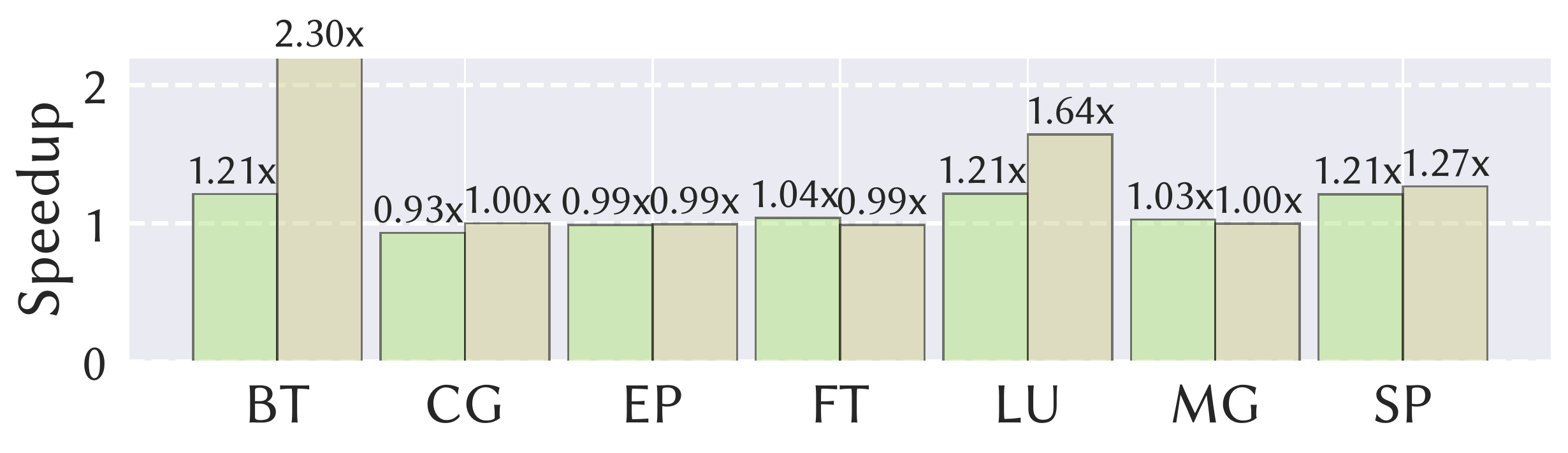}
  \includegraphics[width=0.5\textwidth]{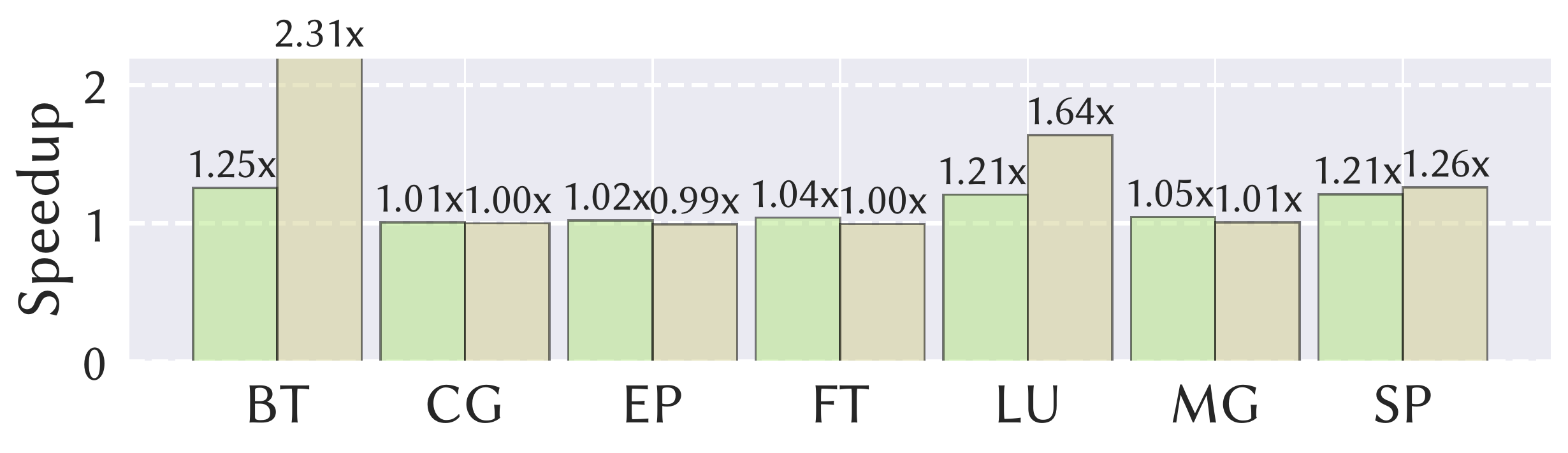}
  }
\begin{tikzpicture}[remember picture,overlay,>=stealth]
  \node[font=\bfseries] at (-0.8,4.8) {\large CSE};
  \node[font=\bfseries] at (-1.5,2) {\large CSE+BULK};
  \node[font=\bfseries] at (7.8,4.8) {\large CSE+SAT};
  \node[font=\bfseries] at (7.85,2) {\large ACCSAT};
\end{tikzpicture}
  \vspace*{-0.35cm}
  \caption{NPB's speedup results on NVIDIA A100-SXM4-80GB. {\color{colornvhpc}$\squadfill$}~NVHPC, {\color{colorgcc}$\squadfill$}~GCC.}
  \label{fig:npb80}
  \vspace*{0.0cm}
\end{figure*}

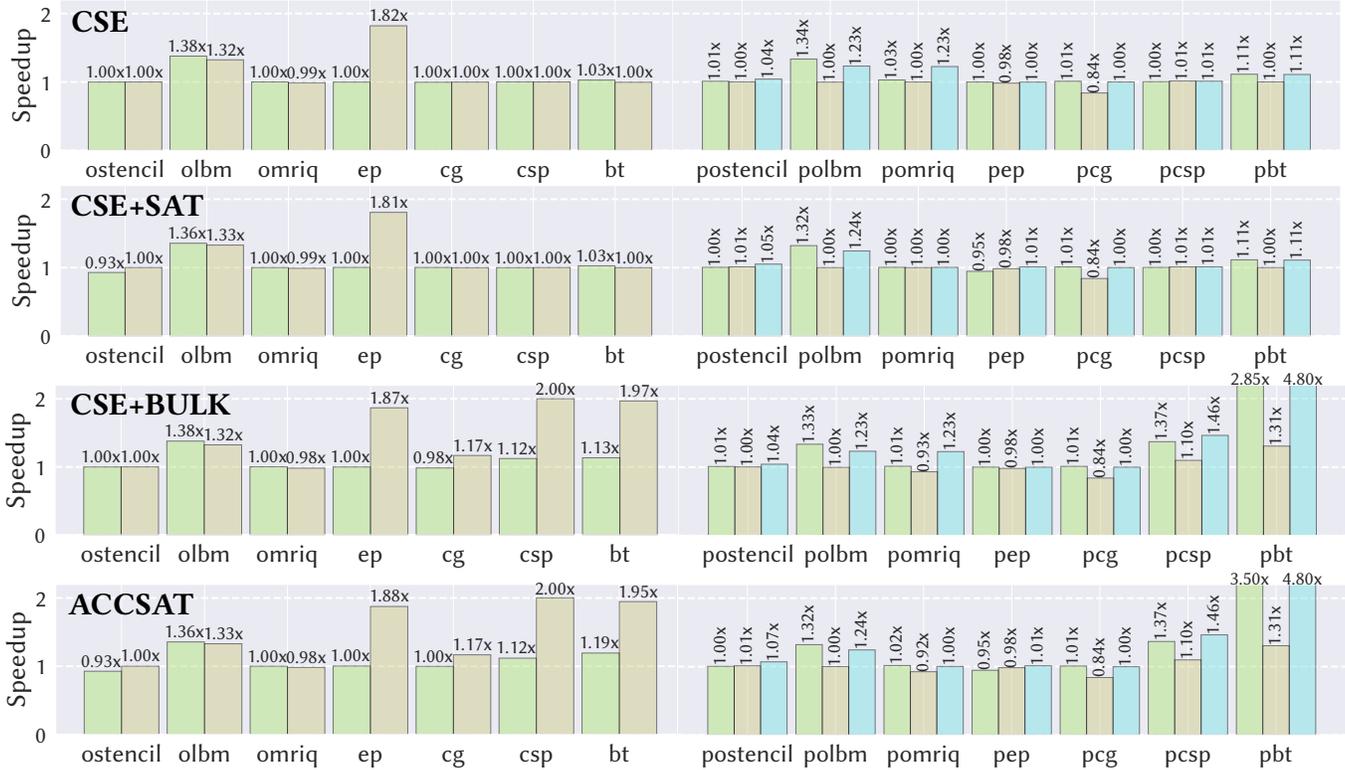
\begin{figure*}[t]
  \vspace*{0.0cm}
  \centering
  \noindent\makebox[\textwidth]{
  \includegraphics[width=0.5\textwidth]{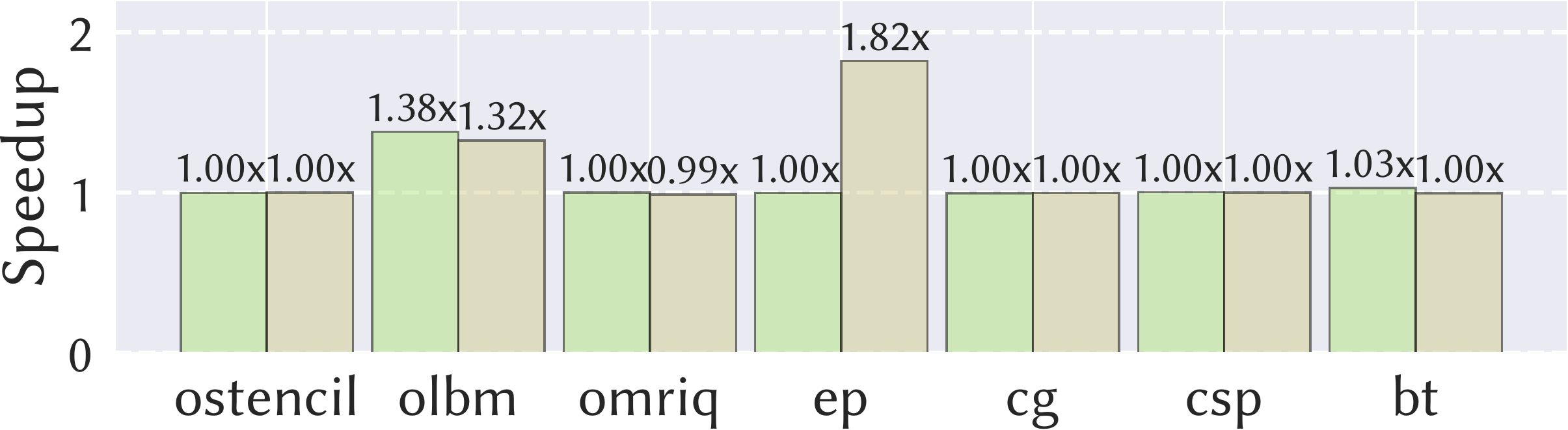}
  \hspace*{-0.8em}
  \includegraphics[width=0.5\textwidth]{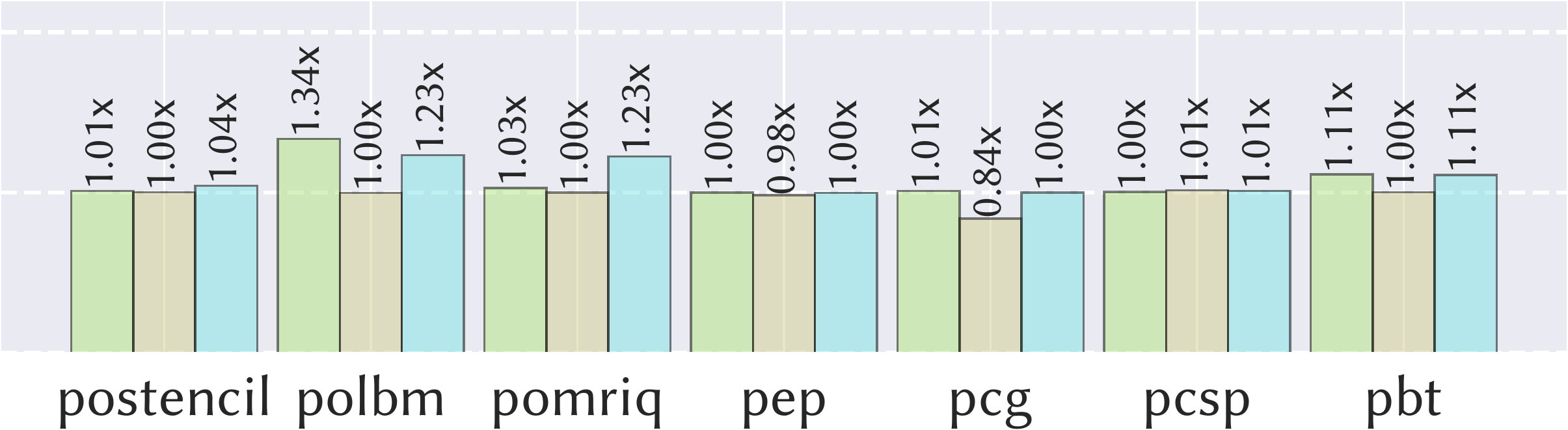}
  }
  \noindent\makebox[\textwidth]{
  \includegraphics[width=0.5\textwidth]{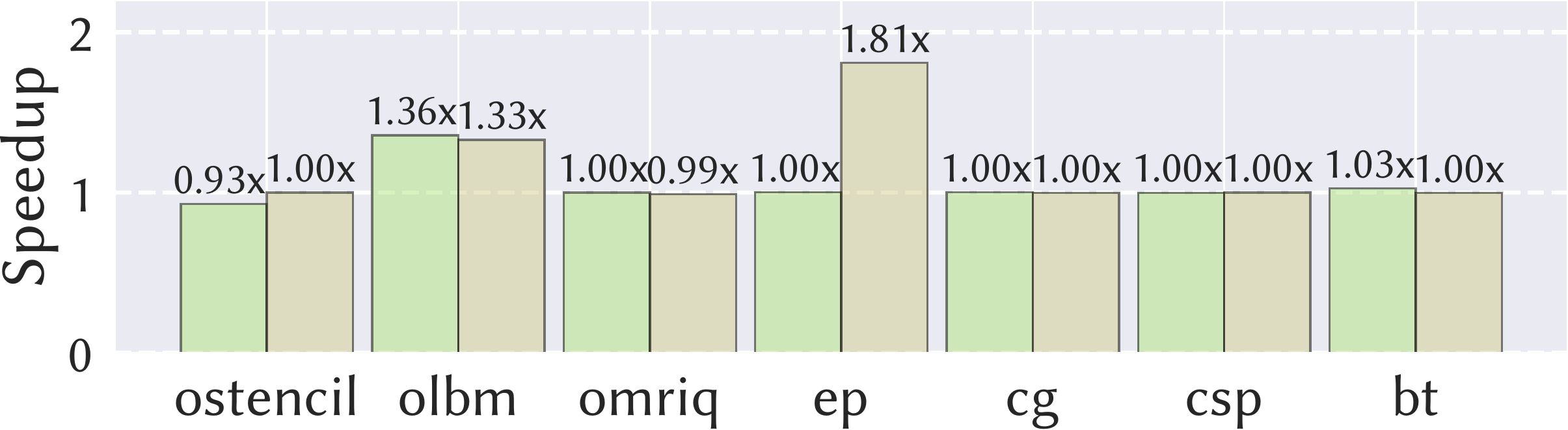}
  \hspace*{-0.8em}
  \includegraphics[width=0.5\textwidth]{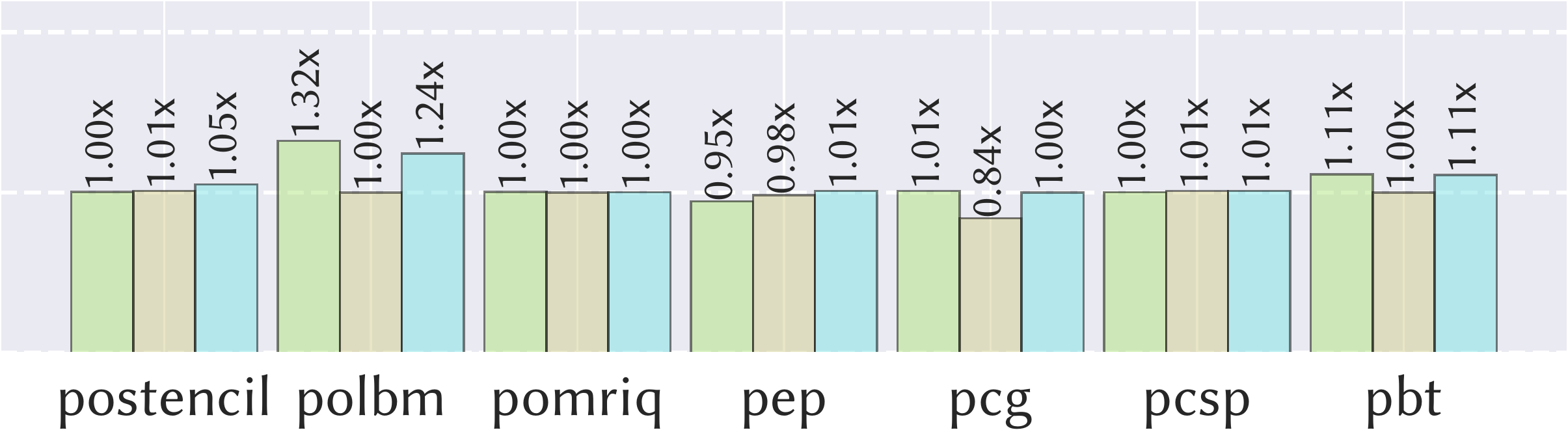}
  }
  \noindent\makebox[\textwidth]{
  \includegraphics[width=0.509\textwidth]{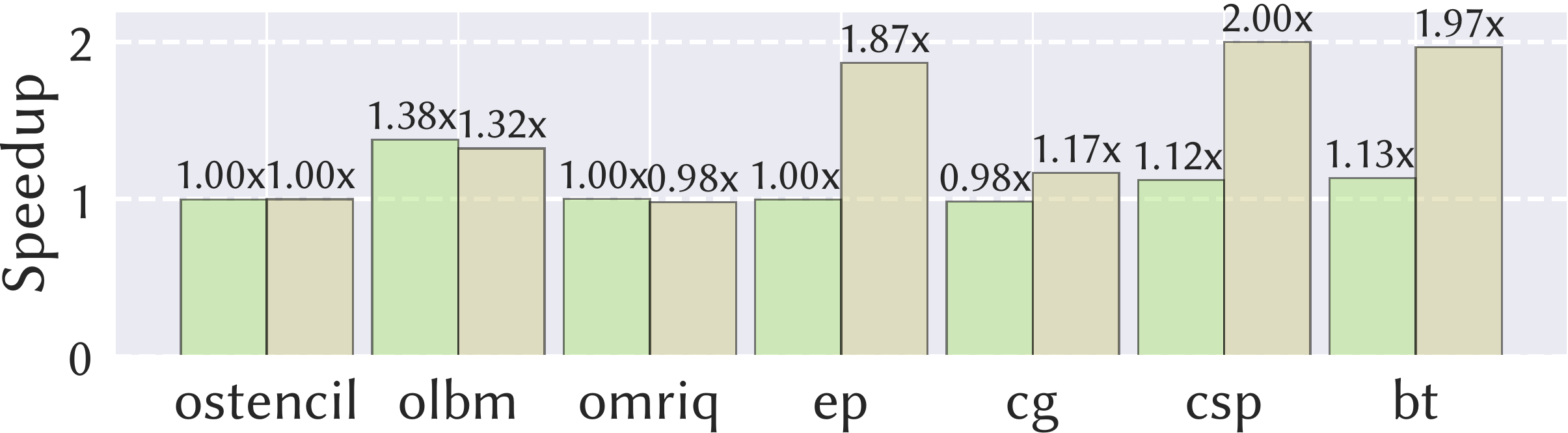}
  \hspace*{-0.8em}
  \includegraphics[width=0.5\textwidth]{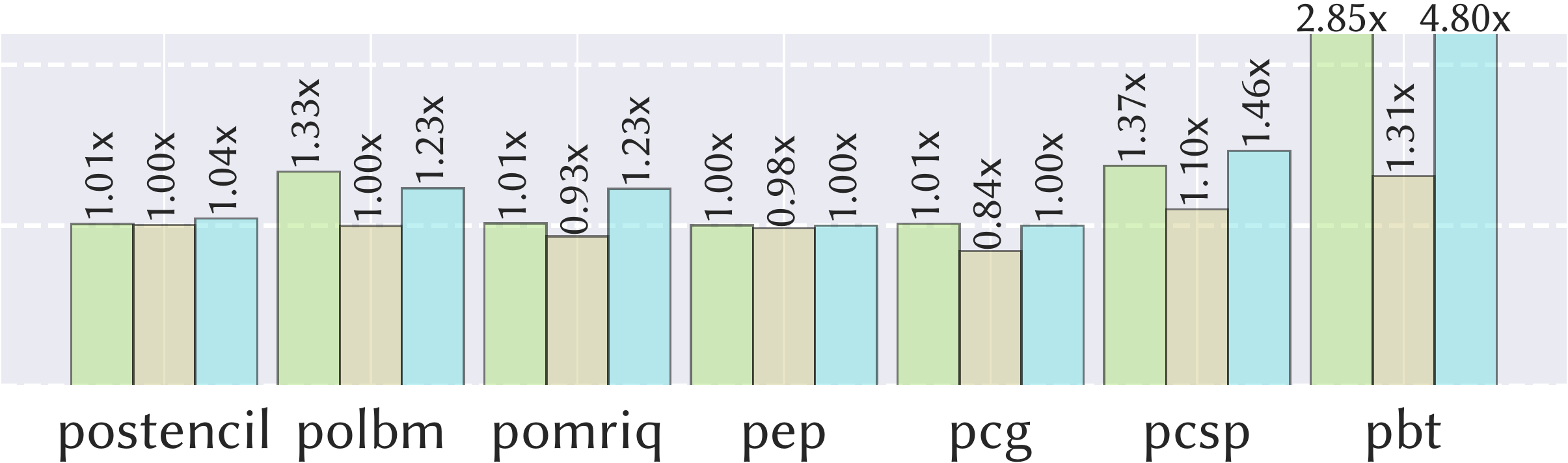}
  }
  \noindent\makebox[\textwidth]{
  \includegraphics[width=0.508\textwidth]{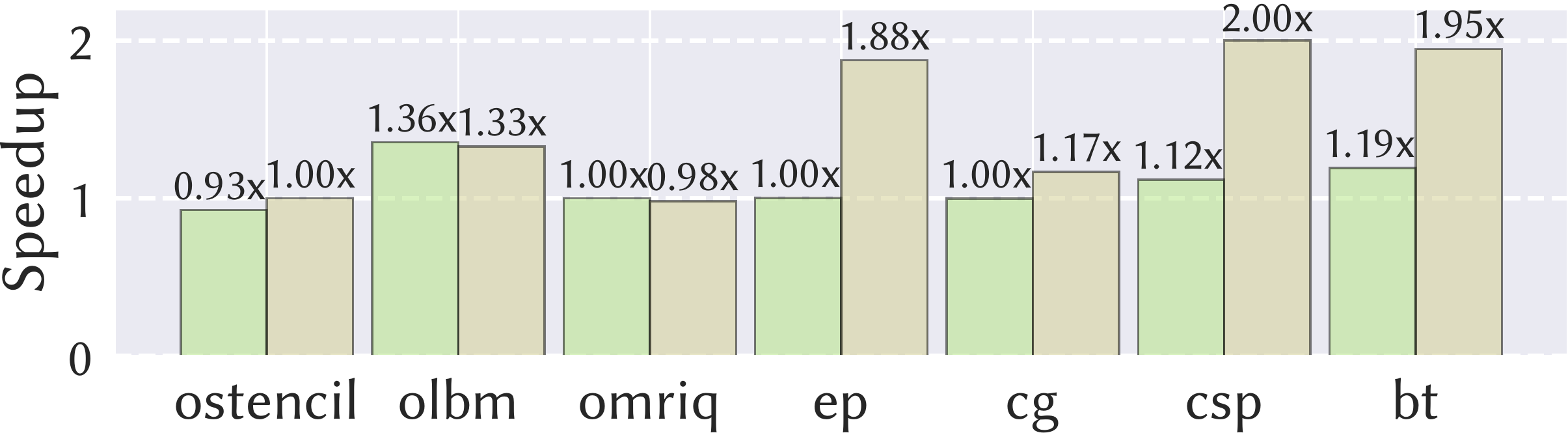}
  \hspace*{-0.8em}
  \includegraphics[width=0.5\textwidth]{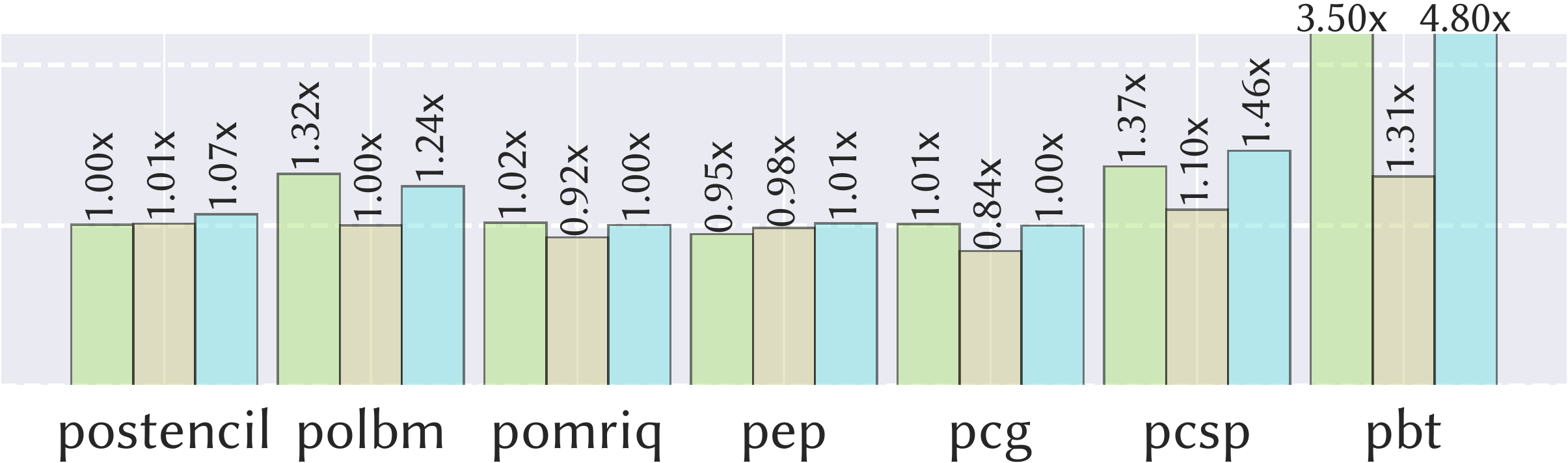}
  }
\begin{tikzpicture}[remember picture,overlay,>=stealth]
  \node[font=\bfseries] at (-7.61,10.35) {\large CSE};
  \node[font=\bfseries] at (-7.12,7.9) {\large CSE+SAT};
  \node[font=\bfseries] at (-6.94,5.25) {\large CSE+BULK};
  \node[font=\bfseries] at (-7.2,2.6) {\large ACCSAT};
\end{tikzpicture}
  \vspace*{-0.7cm}
  \caption{Speedup results of the SPEC ACCEL benchmark suite on NVIDIA A100-SXM4-80GB. {\color{colornvhpc}$\squadfill$}~NVHPC, {\color{colorgcc}$\squadfill$}~GCC, {\color{colorclang}$\squadfill$}~Clang.}
  \label{fig:spec80}
  \vspace*{-0.0cm}
\end{figure*}

Figure~\ref{fig:npb} presents the speedup results of NPB using four generated code versions.
\textbf{CSE} is a version that eliminates redundant loads without performing equality saturation or bulk load.
\textbf{CSE+SAT} and \textbf{CSE+BULK} are \textbf{CSE} with equality saturation and bulk load, respectively.
\textbf{ACCSAT} is a default generated code of ACC Saturator, which includes both equality saturation and bulk load.
With \textbf{ACCSAT}, NVHPC attains up to 1.21x improvement, while GCC attains up to 2.20x speedup.
The \textbf{CSE} version maintains the performance of both compilers,
varying the execution efficiency by 0.98x with NVHPC and by 1.03x with GCC on average.
\textbf{CSE+SAT} provides NVHPC with an average speedup of 0.86\% and improves GCC's performance by only 0.01\%.
However, \textbf{CSE+BULK} significantly accelerates memory-intensive applications such as \textbf{BT}, \textbf{LU}, and \textbf{SP}, while most other benchmarks maintain their performance.
\textbf{ACCSAT} does not degrade the original performance and attains 2.00\% and 0.66\% better throughputs than \textbf{CSE+BULK} on NVHPC and GCC, respectively.
In total, \textbf{ACCSAT} attains average speedups of 1.10x on NVHPC and 1.29x on GCC.

Table~\ref{tab:breakdown} breaks down the top 10 kernels in NPB's \textbf{BT},
and Figure~\ref{fig:breakdown} shows the speedup of each kernel in each version.
The speedups of \textbf{CSE+BULK} and \textbf{ACCSAT} are similar to those of \textbf{CSE} and \textbf{CSE+SAT}, respectively.
\textbf{ACCSAT} attains up to 2.23x and 5.08x improvements on NVHPC and GCC, respectively, resulting in
up to 2.08x and 3.19x memory bandwidth.
The top three kernels on NVHPC suffer performance degradation from \textbf{CSE+BULK} to \textbf{ACCSAT},
because \textbf{ACCSAT} spills more registers to global memory, facilitating the reuse of computation.
The next three kernels execute 8.3\% fewer instructions and achieve around a 1.11x speedup compared to \textbf{CSE+BULK}, since our optimized code clarifies dependencies and reduces both computation and stores.
We expect a total speedup of 1.22x when equality saturation is disabled when registers spill.

On NVHPC, \textbf{CSE} increases the latency of \textbf{CG}, resulting in less SM occupancy due to increased register use. However, \textbf{CSE+SAT} and \textbf{ACCSAT} alleviate this register pressure by utilizing FMA.
Equality saturation enables \textbf{EP} to become faster by executing 10.53\% more FMA operations and 1.67\% fewer total operations than \textbf{CSE}.
Although other parts of \textbf{MG} suffer from register pressure, one part of the benchmark obtains a 1.14x speedup from \textbf{CSE+BULK} to \textbf{ACCSAT} by increasing the L1 cache hit ratio.
Furthermore, \textbf{FT}, \textbf{LU}, and \textbf{SP} benchmarks attain improved memory throughputs and refined total performance with bulk load.

Figure~\ref{fig:spec} depicts the speedups of SPEC benchmarks for both OpenACC and OpenMP versions.
Benchmarks with names starting with "\textbf{p-}" indicate the OpenMP versions.
On average, with OpenACC, \textbf{CSE}, \textbf{CSE+SAT}, \textbf{CSE+BULK}, and \textbf{ACCSAT}
attain 1.06x, 1.04x, 1.08x, and 1.08x speedups, respectively on NVHPC, while with OpenMP, these attain 1.07x, 1.06x, 1.37x, and 1.47x speedups, respectively.
OpenACC's \textbf{ostencil} sees a 16.7\% reduction in SM occupancy due to equality saturation,
leading to decreased performance.
\textbf{CSE} reduces memory loads by around 50\% for \textbf{olbm} and \textbf{polbm}, resulting in improved performance.
Bulk load significantly boosts the performance of memory\-/intensive benchmarks such as \textbf{csp}, \textbf{bt}, \textbf{pcsp}, and \textbf{pbt}, just as in NPB.
\textbf{ACCSAT} attains 3.62x speedup in one part of \textbf{pbt}, which executes only one thread-block over nested loops, by eliminating 77.2\% memory loads and 50.9\% stores. Our optimization decreases operations and reorders memory accesses, being effective for both parallel and sequential iterations.

GCC attains average speedups of 1.16x, 1.16x, 1.48x, and 1.48x for OpenACC using \textbf{CSE}, \textbf{CSE+SAT}, \textbf{CSE+BULK}, and \textbf{ACCSAT}, respectively, and 1.00x, 1.00x, 1.06x, and 1.06x for OpenMP.
The original versions of OpenMP result in high register pressure, which limits the benefits of bulk load. Conversely, the initial versions of OpenACC use fewer registers while setting inadequate parallelism, likely due to the immature support of OpenACC's kernels directive.
\textbf{CSE} reduces memory loads by 54.8\% in \textbf{olbm}, yielding a 1.32x speedup.
For \textbf{ep}, \textbf{CSE} minimizes operations, and bulk load enhances overall memory throughput.
Bulk load also benefits \textbf{cg}, \textbf{csp}, \textbf{bt}, \textbf{pscp}, and \textbf{pbt} by mitigating global-memory latency.
However, \textbf{pomriq}'s SM occupancy decreases with bulk load and equality saturation, leading to reduced efficiency.

Clang attains average speedups of 1.06x, 1.06x, 1.69x, and 1.66x for OpenMP using \textbf{CSE}, \textbf{CSE+SAT}, \textbf{CSE+BULK}, and \textbf{ACCSAT}, respectively.
\textbf{CSE+SAT} refines the performance of \textbf{postencil}, \textbf{polbm}, and \textbf{pep} compared to \textbf{CSE}.
\textbf{ACCSAT} further improves \textbf{pbt} over \textbf{CSE+BULK}, while reducing \textbf{pomriq}'s ILP due to decreased register usage. \textbf{CSE+BULK} attains a maximum speedup of 4.84x through optimized memory accesses.

For comparison, Figure~\ref{fig:npb80} provides the speedups of NPB on
an NVIDIA A100-SXM4-80GB, which features 1.31x higher memory bandwidth
than the A100-PCIE-40GB.
The original performance is improved by 5.79\% with NVHPC and 4.65\% with GCC on average,
while most benchmarks preserve the performance changes on the other GPU.
We confirm that \textbf{BT} gains further 1.25x and 2.31x speedups by \textbf{ACCSAT} on NVHPC and GCC, respectively.
Using the same GPU, \textbf{CSE+BULK} improves the execution by 1.21x and 2.30x,
and \textbf{ACCSAT} improves both computation and memory throughputs in \textbf{BT} more for faster memory systems as the latency of computation becomes distinct.
With \textbf{ACCSAT}, NVHPC and GCC attain average speedups of 1.11x and 1.31x, respectively.
Our technique mitigates the memory latency that results from GCC's inadequate thread utilization for the kernels directive, resulting in superior performance gains compared to NVHPC.

Figure~\ref{fig:spec80} shows the speedups of SPEC benchmarks on NVIDIA A100-SXM4-80GB.
With OpenACC, NVHPC increases the original performance by 7.42\% and GCC increases by 3.11\% using the GPU on average,
while the original performance with OpenMP is increased by 3.33\% on NVHPC, -13.3\% on GCC, and 1.04\% on Clang.
Especially, \textbf{pcg} on GCC suffers from the latency of memory barriers, which decreases the original performance by 59.3\% and degenerates the execution of optimized codes.
\textbf{CSE+SAT} causes a lower L1 cache hit ratio for \textbf{pep} on NVHPC.
Overall, \textbf{ACCSAT} obtains average speedups of 1.09x on NVHPC and 1.47x on GCC with OpenACC and average speedups of 1.45x on NVHPC, 1.02x on GCC, and 1.66x on Clang with OpenMP.

\section{Related Work}\label{sec:related}

Maximizing architectural utilization is a significant challenge in HPC, requiring application code optimization. Compilers play a crucial role for enhancing performance by applying various code-generation techniques, enabling both generic and architecture-specific optimizations. Numerous programming models and state-of-the-art techniques, especially for GPUs, have been introduced to adapt codes to parallel architectures~\cite{halide, cuda, docon, pytorch, kokkos}.
Directive-based programming models~\cite{openacc,openmp} extend sequential languages, allowing complex scientific applications to offload loop iterations to accelerators while maintaining their structures. However, such compilers often rely on the basis of sequential code generation, thus limiting optimization opportunities to general computation~\cite{jacc, barua}.

Several projects have explored domain-specific or
architecture\-/specific approaches for directive-based code. The CLAW
DSL~\cite{claw} provides directives for grid-based algorithms,
enabling target-specific optimizations such as spatial blocking while
supporting OpenACC/OpenMP code generation. JACC~\cite{jacc} is an
OpenACC runtime framework offering just-in-time kernel compilation
with dynamic constant expansion. OptACC~\cite{autotuningacc} performs
runtime parameter searches to optimize OpenACC
parallelism. CCAMP~\cite{ccamp} interchanges OpenACC and OpenMP,
optimizing parallelization for each combination of models and
architecture. Barua {\it et al.}~\cite{barua} develop an automated OpenACC-kernel optimizer for maximizing ILP through loop unfolding. SAFARA~\cite{daniel} fully utilizes register resources to facilitate array reference reuse in OpenACC kernels.

Our ACC Saturator differs in three aspects: (1) automation of optimization through rewriting rules and a cost model-based optimal code selection, (2) integration of bulk load optimization technique for significant GPU memory throughput improvement, and (3) preservation of original code structures while being applicable to both OpenACC and OpenMP without requiring domain-specific information.

Since the introduction of the equality-saturation library
egg~\cite{egg}, numerous studies have leveraged it, particularly in
the context of GPU computing, for accelerating deep learning
applications~\cite{MLSYS2021_65ded535, 10.1145/3559009.3569673,
  shaikhha2023nablasd, 10.14778/3407790.3407799,
  10.1145/3460945.3464953}. These works employ rewriting rules for
arithmetic expressions, abstract operations, or tensor graphs to
optimize convolutions, sparse tensor algebra, or whole tensor
operations. Diospyros~\cite{dsp} synthesizes efficient DSP operations
from C code using equality saturation, while Gowda {\it et al.}~\cite{10.1145/3511528.3511535} implement a symbolic algebra system with egg for automatic parallelism assignment.

Although the initial work of equality saturation was demonstrated as a Java bytecode optimizer~\cite{equality}, recent works using egg focus on program synthesis and code optimization without control statements. Our work is the first to bridge the gap between user code and equality saturation by automatically extracting computation and constructing SSA information for data dependencies. This approach enables novel equality\-/saturation optimizations for directive-based programming, as detailed in this paper, without requiring further programmer intervention.

Several works propose innovative code optimization approaches. Ben-Nun
{\it et al.}~\cite{10.1145/3295500.3356173, mlir-dace} develop
data-centric flow graphs to focus on data-oriented optimizations, such
as dead memory elimination. Their representation supports collective
operations and macroscopic parallelization, enabling optimizations
irrespective of calculation. The MLIR framework~\cite{multiir}
utilizes multi-level IRs for cooperative domain-specific
optimizations. Ginsbach {\it et al.}~\cite{10.1145/3296957.3173182} define code patterns for performance opportunities, replacing matched code with library or DSL implementations. Additionally, machine learning is gaining popularity for automatic compiler-optimization tuning~\cite{10.1109/CGO53902.2022.9741263, trofin2021mlgo, 10.1145/3368826.3377928}. Our tool employs e-graph operations to identify optimal solutions without source code analysis or abstract syntax tree transformations.

\section{Conclusion}\label{sec:conclusion}

In recent years, there has been a surge in the development of
automation tools to maximize the potential of GPUs for various
applications. Directive-based programming, in particular, has become a
popular method for converting sequential code into parallel code
through annotations. However, these abstract models may hinder code
optimization efforts, since actual applications often have complex data dependencies that cannot be easily extended within sequential code structures. As a result, techniques for targeting sequential code to GPUs have been developed to attain higher throughputs while accommodating the unique features of GPU parallel architectures.

This paper presents ACC Saturator, an equality\-/saturation framework for directive-based programming models that optimizes code using rewriting rules and a cost model. It is the first framework to bridge the gap between user code and equality saturation optimizations while preserving original code structures and data dependencies. We introduce a novel technique, bulk load, which we enable through our framework, to generate code with intentional high memory pressure. We demonstrate the effectiveness of ACC Saturator on various practical benchmarks using multiple compilers for both OpenACC and OpenMP on a state-of-the-art GPU architecture. Our analysis highlights the significance of memory-access order and computational reordering, which ACC Saturator enables, for achieving significant performance improvements and increased memory throughput.

\section*{Acknowledgement}

The authors are funded by the EPEEC project from the European Union's Horizon 2020 research and innovation program under grant agreement No. 801051, the Ministerio de Ciencia e Innovación—Agencia Estatal de Investigación (PID2019-107255GB-C21/\-AEI/10.13039/501100011033), and 
Departament de Recerca i Universitats from the Generalitat de Catalunya as the Research Group AccMem (Code: 2021 SGR 00807).
Antonio J. Pe\~{n}a was partially supported by the Ram\'on y Cajal fellowship RYC2020-030054-I funded by MCIN/AEI/10.13039/501100011033 and by “ESF Investing in your future”.
The authors gratefully acknowledge the support of the NVIDIA Solutions Lab who provided us the remote access to their GPU environment.
The authors would like to acknowledge the NVIDIA AI Technology Center (NVAITC) Europe for their valuable help. Sandia National Laboratories is a multimission laboratory managed and operated by National Technology \& Engineering Solutions of Sandia, LLC, a wholly owned subsidiary of Honeywell International Inc., for the U.S. Department of Energy’s National Nuclear Security Administration under contract DE-NA0003525.

\balance
\bibliographystyle{IEEEtran}
\bibliography{main}

\end{document}